\documentclass[pre,showpacs,showkeys,preprintnumbers,amsmath,amssymb,superscriptaddress,twocolumn]{revtex4}
\usepackage{graphicx}
\usepackage{amsfonts}
\usepackage{url}
\usepackage{epstopdf}
\usepackage{color}
\usepackage{bm}
\usepackage[colorlinks=true,linkcolor=blue,citecolor=red]{hyperref}%

\def\L{\mathcal L}
\def\N{\mathcal N}

\def\n{\bm{n}}
\def\s{\bm{s}}
\def\x{\bm{x}}
\def\X{\bm{X}}

\def\pa{\partial\Omega}

\def\I{{\mathbb I}}

\def\R{{\mathbb R}}
\def\Z{{\mathbb Z}}

\def\L{{\mathcal L}}

\def\M{{\mathcal M}}

\def\y{\bm{y}}
\def\X{\bm{X}}

\def\erf{\mathrm{erf}}
\def\erfc{\mathrm{erfc}}
\def\erfcx{\mathrm{erfcx}}
\def\ctanh{\mathrm{ctanh}}

\begin{document}

\title{Surface Hopping Propagator: \\ An Alternative Approach to Diffusion-Influenced Reactions}

\author{Denis~S.~Grebenkov}
 \email{denis.grebenkov@polytechnique.edu}
\affiliation{
Laboratoire de Physique de la Mati\`{e}re Condens\'{e}e (UMR 7643), \\ 
CNRS -- Ecole Polytechnique, IP Paris, 91128 Palaiseau, France}

\date{\today}

\begin{abstract}
Dynamics of a particle diffusing in a confinement can be seen a
sequence of bulk-diffusion-mediated hops on the confinement surface.
Here, we investigate the surface hopping propagator that describes the
position of the diffusing particle after a prescribed number of
encounters with that surface.  This quantity plays the central role in
diffusion-influenced reactions and determines their most common
characteristics such as the propagator, the first-passage time
distribution, and the reaction rate.  We derive explicit formulas for
the surface hopping propagator and related quantities for several
Euclidean domains: half-space, circular annuli, circular cylinders,
and spherical shells.  These results provide the theoretical ground
for studying diffusion-mediated surface phenomena.  The behavior of
the surface hopping propagator is investigated for both ``immortal''
and ``mortal'' particles.
\end{abstract}


%02.50.-r       (Probability theory, stochastic processes, and statistics)
%05.40.-a 	Fluctuation phenomena, random processes, noise, and Brownian motion
%02.70.Rr       (General statistical methods)
%05.10.Gg 	Stochastic analysis methods (Fokker-Planck, Langevin, etc.) 

%02.50.Ey 	Stochastic processes  (Probability theory, stochastic processes, and statistics)


\maketitle

\section{Introduction}

In many natural phenomena, particles diffuse in a confinement towards
its surface where they can react, permeate, relax their activity or be
killed.  Examples include heterogeneous catalysis, permeation across
cell membranes, filtering in porous media, surface relaxation in
nuclear magnetic resonance, and animal foraging
\cite{Rice,Redner,Schuss,Metzler,Oshanin,Grebenkov07,Benichou11,Bressloff13,Benichou14}.
These phenomena are conventionally described by diffusion equation (or
more general Fokker-Planck equation) with appropriate boundary
conditions \cite{Gardiner,Risken}.  In particular, most common
properties of diffusion-influenced reactions are derived from the
propagator $G_q(\x,t|\x_0)$, i.e., the probability density of the
event that a particle, started from a bulk point $\x_0$ at time $0$,
has not reacted on the surface and located at a bulk point $\x$ at
time $t$.  For normal diffusion, the propagator satisfies the
diffusion equation inside a confining domain $\Omega$
\begin{equation}  \label{eq:diff_eq}
\partial_t G_q(\x,t|\x_0) = D \, \Delta_{\x} G_q(\x,t|\x_0) \quad (\x\in\Omega), 
\end{equation}
subject to the initial condition $G_q(\x,t=0|\x_0) = \delta(\x-\x_0)$
and the Robin boundary condition on the boundary $\pa$:
\begin{equation}  \label{eq:Robin}
- \partial_{\n} G_q(\x,t|\x_0) = q \, G_q(\x,t|\x_0) \quad (\x\in\pa),
\end{equation}
where $\Delta_{\x}$ is the Laplace operator acting on $\x$,
$\delta(\x)$ is the Dirac distribution, and $\partial_{\n}$ is the
normal derivative on the boundary $\pa$ oriented outwards the domain
$\Omega$.
The parameter $q = \kappa/D$ is the ratio between the surface
reactivity (or permeability, or relaxivity, etc.) $\kappa$ and bulk
diffusivity $D$.  In chemical physics, the Robin boundary condition
was put forward by Collins and Kimball \cite{Collins49} and later
explored by many researchers
\cite{Lauffenburger,Sano79,Sano81,Shoup82,Zwanzig90,Sapoval94,Filoche99,Sapoval02,Grebenkov03,Berezhkovskii04,Grebenkov05,Grebenkov06a,Traytak07,Bressloff08,Lawley15,Galanti16,Lindsay17,Grebenkov17,Bernoff18b,Grebenkov19d}
(see an overview in \cite{Grebenkov19b}).  The major disadvantage of
the conventional description is that the surface reactivity $\kappa$
(or $q$) enters {\it implicitly} as a parameter of the Robin boundary
condition (\ref{eq:Robin}).

In a recent work \cite{Grebenkov20}, we proposed an alternative
description based on the concept of boundary local time.  The boundary
local time $\ell_t$ characterizes the fraction of time that a
diffusing particle spends in a close vicinity of the reflecting
boundary, as well as the number of encounters with that boundary
\cite{Levy}, see Eqs. (\ref{eq:ellt1}, \ref{eq:ellt2}) below.  This is
a fundamental concept in the theory of stochastic processes
\cite{Ito,Freidlin}, which remains largely unknown and almost
unemployed in physics, chemistry and biology.  To incorporate
$\ell_t$, we introduced the full propagator $P(\x,\ell,t|\x_0)$, i.e.,
the joint probability density of finding a particle at point $\x$ at
time $t$ with its boundary local time $\ell$, given that it started
from $\x_0$ at time $0$.  The crucial advantage of this alternative
description is that $P(\x,\ell,t|\x_0)$ characterizes diffusion in
confinement with {\it reflecting} (inert) boundary.  In turn, the
surface reactivity is introduced via a stopping condition on the
boundary local time.  In particular, we derived
\begin{equation}  \label{eq:Gq_P}
G_q(\x,t|\x_0) = \int\limits_0^\infty d\ell \, e^{-q\ell} \, P(\x,\ell,t|\x_0) ,
\end{equation}
where the surface reactivity $q$ appears {\it explicitly} as a
parameter of the Laplace transform with respect to the boundary local
time $\ell$.  In this way, the single full propagator
$P(\x,\ell,t|\x_0)$ describes the whole family of partially reactive
surfaces (characterized by $q$).  Moreover, one can replace the
exponential factor $e^{-q\ell}$ by a more general function to
implement other surface reaction mechanisms far beyond the
conventional partial reactivity described by the Robin boundary
condition (\ref{eq:Robin}), see
\cite{Grebenkov20} for details.  In this light, the full propagator
$P(\x,\ell,t|\x_0)$ turns out to be the intrinsic key quantity that
describes all sorts of diffusion-mediated surface phenomena in a given
confinement.

A successful implementation of this new paradigm requires efficient
methods for accessing the full propagator.  In \cite{Grebenkov20}, the
Laplace transform of the full propagator was expressed in terms of the
so-called ``surface hopping propagator'' $\Sigma_p(\s,\ell|\s_0)$,
i.e., the probability density of the event that a particle, started
from a boundary point $\s_0$, has survived against a ``bulk killing''
with the rate $p$ and located at a boundary point $\s$ at the boundary
local time $\ell$.  The rate $p \geq 0$ accounts for eventual
disappearance of the particles during its diffusion in the domain
$\Omega$ due to a bulk reaction or spontaneous disintegration,
relaxation, photobleaching or death.  In this scheme, one can consider
both ``mortal'' ($p >0$) and ``immortal'' ($p = 0$) particles
\cite{Yuste13,Meerson15,Grebenkov17d}.  In other words, the surface
hopping propagator describes bulk-diffusion-mediated displacements
between two encounters with the boundary, separated by the boundary
local time $\ell$.  The concept of such a surface exploration by
successive hops through the bulk was formulated by Bychuk and
O'Shaugnessy \cite{Bychuk94,Bychuk95} and later confirmed by
single-particle tracking experiments \cite{Walder11,Skaug13,Wang17}.
Former theoretical descriptions of surface hopping diffusion in terms
of effective surface propagators were based on coupled bulk-surface
diffusion equations with adsorption/desorption kinetics
\cite{Chechkin09,Chechkin11,Chechkin12,Berezhkovskii15,Berezhkovskii17}.
In turn, the surface hopping propagator $\Sigma_p(\s,\ell|\s_0)$ is a
conceptually different quantity, which characterizes surface
displacements not in terms of {\it physical time} $t$ (as earlier) but
in terms of the boundary local time $\ell$ (the number of encounters).
To our knowledge, the surface hopping propagator, introduced in
\cite{Grebenkov20} as an efficient way to access the full propagator,
is a new object, and the present paper aims at uncovering its
properties.

The paper is organized as follows.  In Sec. \ref{sec:theory}, we
formulate the theoretical framework for diffusion-mediated surface
phenomena, build an intuitive ground for the surface hopping
propagator, and recall some general relations from \cite{Grebenkov20}.
Main results are reported in Sec. \ref{sec:examples}, in which the
surface hopping propagator is computed and investigated for several
domains.  Section \ref{sec:discussion} summarizes and concludes the
paper.

\section{Surface hopping propagator}
\label{sec:theory}

How many reflections does a particle undertake up to a given time $t$
or during its lifetime?  Where is the particle after $n$ reflections?
For the common continuous-time Brownian motion, these natural
questions have old but disappointing (trivial) answers.  In fact,
Brownian motion crossing a smooth surface is known to return
infinitely many times to that surface within an infinitely short time
period \cite{Morters}.  To get more satisfactory answers, one needs to
reformulate these questions in a regularized way.  For instance, one
can substitute Brownian motion by a sequence of independent jumps
(e.g., a random walk on a lattice).  However, it is more convenient to
keep considering continuous stochastic process $\X_t$ but to introduce
a thin surface layer of width $a$, $\pa_a = \{ \x \in \Omega ~:~ |\x -
\pa| < a\}$, and to count the number $\N_t^a$ of crossings of this
layer by reflected Brownian motion up to time $t$.  As $a \to 0$, the
number of crossings diverges but $a\, \N_t^a$ converges to the random
process, introduced by L\'evy and called the boundary local time
\cite{Levy}:
\begin{equation}  \label{eq:ellt1}
\ell_t = \lim\limits_{a\to 0} a\, \N_t^a .
\end{equation}
While the continuous time $t$ represents the number of jumps of
duration $\delta$ that the particle undertakes in the bulk, the
boundary local time $\ell$ is the proxy of the number of encounters
with the boundary (reflections of amplitude $a$).  Equivalently,
$\ell_t$ is proportional to the fraction of time that a particle spent
in the surface layer of width $a$ up to time $t$:
\begin{equation}   \label{eq:ellt2}
\ell_t = \lim\limits_{a\to 0} \frac{D}{a} \int\limits_0^t dt' \, \I_{\pa_a}(\X_{t'}) ,
\end{equation}
where the integral is the residence time of reflected Brownian motion
$\X_t$ in $\pa_a$, and $\I_{\pa_a}(\x)$ is the indicator function of
that layer: $\I_{\pa_a}(\x) = 1$ for $\x \in \pa_a$, and $0$
otherwise.  While $\ell_t$ is historically called ``local time'', it
has units of length (see also \cite{Grebenkov07a,Grebenkov19c}).
In some definitions, the diffusion coefficient $D$ is removed
from Eq. (\ref{eq:ellt2}), yielding the boundary local time in units
of time per length, i.e., the time spent in the surface layer rescaled
by its width.  We also stress that the boundary local time should not
be confused with a closely related notion of the point local time,
i.e., a fraction of time spent in an infinitesimal vicinity of a fixed
bulk point.  The latter was thoroughly investigated, in particular,
for Brownian motion and Bessel processes (see
\cite{Borodin,Takacs95,Randon18} and references therein).

The former two questions should thus be reformulated in terms of the
boundary local time: How large is the boundary local time $\ell_t$ up
to a given time $t$ or during the lifetime of a particle?  Where is
the particle after a boundary local time $\ell$?  Answers to both
these questions are given by the surface hopping propagator
$\Sigma_p(\s,\ell|\s_0)$, as discussed below.

\subsection{Intuitive picture}

\begin{figure}
\begin{center}
\includegraphics[width=85mm]{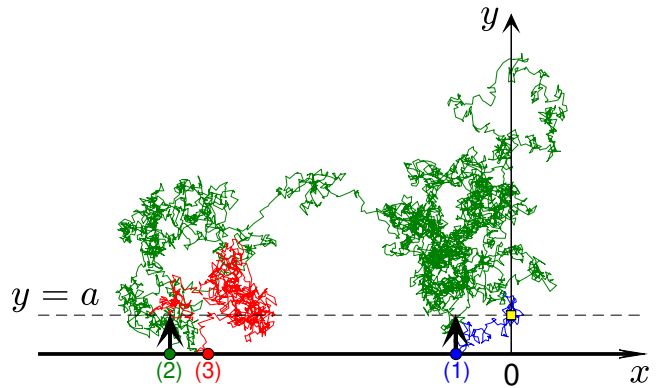} 
\end{center}
\caption{
Simulated trajectory of Brownian motion in the upper half-plane above
the horizontal axis.  A thin surface layer is delimited by dashed line
at $y = a$.  The trajectory, started from $(0,a)$ (yellow filled
square) is split into three colored parts (blue, green, red).  Each
part is terminated when the particle hits the boundary (enumerated
filled circles), while the next part starts the distance $a$ above the
last hitting point (``jumps'' indicated by black arrows).  While some
bulk explorations are short (blue and red parts), the other can be
very long (green part).}
\label{fig:scheme}
% [X,Y,Ir] = A_localtime_BM_fig4;
% 'BM_half1.mat'
\end{figure}

Before presenting main results for general domains, it is instructive
to provide the motivation and intuition for the surface hopping
propagator.  Let us consider a particle diffusing in the upper
half-plane, $\Omega = \R \times \R_+$, with reflecting horizontal axis
$\pa = \{(x,0) ~:~ x\in \R\}$.  If the particle started from a
boundary point $\s_0 = (x_0,0)$, its ``next'' encounter with the
boundary would occur exactly at $\s_0$, as discussed above.  To
overcome this problem, we introduce a thin surface layer of width $a$
(Fig. \ref{fig:scheme}).  Now, one can ask what is the position of the
next encounter with the boundary after crossing the horizontal line $y
= a$ (i.e., after exiting from the surface layer).  This
regularization eliminates too short Brownian trajectories that remain
within the layer.  As the first crossing of the line $y = a$ typically
occurs near the starting point $\s_0$, one can move the starting point
from $(x_0,0)$ to $(x_0,a)$ and then search for the probability
density of the first arrival onto the horizontal axis.  This is the
harmonic measure density, which for the upper half-plane takes the
form of the Cauchy density,
\begin{equation}
p_1(x|(x_0,a)) = \frac{a}{\pi [(x-x_0)^2 + a^2]} \,,
\end{equation}
and can thus describe the first encounter position $x$ after leaving
the boundary from $x_0$ and crossing the surface layer of width $a$.
After this encounter, the particle continues diffusion, independently
of its past, so that the second encounter position is determined by
the convolution of two Cauchy densities:
\begin{align*}  \nonumber
p_2(x|(x_0,a)) & = \int\limits_{\R} dx_1\, p_1(x|(x_1,a)) \, p_1(x_1|(x_0,a)) \\
& = \frac{2a}{\pi [(x-x_0)^2 + (2a)^2]}  \,.
\end{align*}
Similarly, the position of the $n$-th encounter is determined by
\begin{equation}  \label{eq:pn}
p_n(x|(x_0,a)) = \frac{na}{\pi [(x-x_0)^2 + (na)^2]} \,.
\end{equation}
In the limit $a \to 0$ with any fixed $n$, this density converges to
the Dirac distribution, $p_n(x|(x_0,a)) \to \delta(x-x_0)$,
illustrating the above statement that (reflected) Brownian motion
returns infinitely many times to the first hitting point within an
infinitely short period.  As the right-hand side of Eq. (\ref{eq:pn})
depends on $a$ via the product $na$, a nontrivial result can only be
obtained in the limit $a\to 0$ when $na$ is fixed.  Setting $\ell =
na$, one obtains the surface hopping propagator for the upper
half-plane:
\begin{equation} \label{eq:Cauchy}
\Sigma_0(\s,\ell|\s_0) = \frac{\ell}{\pi [(x-x_0)^2 + \ell^2]} \,,
\end{equation}
with boundary points $\s = (x,0)$ and $\s_0 = (x_0,0)$.  As eventual
death of the particle during its bulk diffusion was ignored, we set $p
= 0$ in the subscript.

While the above construction can be performed in any confining domain,
its practical realization involves numerous convolutions of the
harmonic measure density which in general are difficult to compute
(the above explicit computation was possible due to the explicit form
of the Cauchy distribution and its specific ``infinite divisibility''
property, i.e., the invariance of its form upon convolutions).  In the
next subsection, we present a general approach to access the surface
hopping propagator.

\subsection{General approach}

We consider a particle diffusing in an Euclidean domain $\Omega\subset
\R^d$ with a smooth boundary $\pa$.  In \cite{Grebenkov20}, the
surface hopping propagator $\Sigma_p(\s,\ell|\s_0)$ (with $p \geq 0$)
was shown to be the kernel of the semi-group $\exp(-\M_p \ell)$
generated by the Dirichlet-to-Neumann operator $\M_p$.  This is a
pseudo-differential self-adjoint operator acting on functions on the
boundary $\pa$ (see rigorous definitions and mathematical details in
\cite{Arendt14,Daners14,Arendt15,Hassell17,Girouard17}).  For a given
function $f$ on $\pa$, this operator associates another function $g$
on $\pa$, $\M_p ~:~ f\to g = (\partial_{\n} w)_{|\pa}$, where $w$
satisfies the Dirichlet boundary value problem:
\begin{equation}  \label{eq:u_def}
(p - D \Delta) w = 0 \quad \textrm{in}~\Omega, \qquad w_{|\pa} = f 
\end{equation}
(if $\Omega$ is unbounded, one also needs to impose the regularity
condition: $w\to 0$ as $|\x|\to \infty$, see below).  For instance, if
$f$ describes a concentration of particles maintained on the boundary
$\pa$, then $\M_p f = (\partial_{\n} w)_{|\pa}$ is proportional to the
diffusive flux of these particles into the bulk.  Note that there is a
family of Dirichlet-to-Neumann operators parameterized by $p$ (or
$p/D$).  As the kernel of the semi-group $\exp(-\M_p\ell)$, the
surface hopping propagator satisfies
\begin{equation}
\partial_\ell \Sigma_p(\s,\ell|\s_0) = - \M_p \Sigma_p(\s,\ell|\s_0),
\end{equation}
subject to the initial condition $\Sigma_p(\s,\ell=0|\s_0) = \delta(\s
- \s_0)$.  This equation resembles the diffusion equation
(\ref{eq:diff_eq}), in which the physical time $t$ is replaced by the
boundary local time $\ell$, and the Laplace operator $\Delta$ is
replaced by $-\M_p$.

When the boundary $\pa$ of the domain is bounded, the
Dirichlet-to-Neumann operator has a discrete spectrum, i.e., a
countable set of positive eigenvalues $\mu_n^{(p)}$ and eigenfunctions
$v_n^{(p)}(\s)$ forming a complete orthonormal basis in the space
$L_2(\pa)$:
\begin{equation}  \label{eq:M_eigen}
\M_p \, v_n^{(p)}(\s) =  \mu_n^{(p)} \, v_n^{(p)}(\s) .
\end{equation}
The surface hopping propagator admits thus the spectral expansion:
\begin{equation}  \label{eq:Sigma_M}
\Sigma_p(\s,\ell|\s_0) = \sum\limits_n [v_n^{(p)}(\s_0)]^* \, v_n^{(p)}(\s) \, e^{-\mu_n^{(p)} \ell} ,
\end{equation}
where asterisk denotes complex conjugate.  In other words, finding
$\Sigma_p(\s,\ell|\s_0)$ is equivalent to studying the spectral
properties of the Dirichlet-to-Neumann operator $\M_p$.

When $p > 0$, all eigenvalues $\mu_n^{(p)}$ are strictly positive, and
the surface hopping propagator vanishes as $\ell$ increases.  This is
a direct consequence of bulk reaction that may lead to eventual death
or disappearance of the diffusing particle during its motion.  In
particular, 
\begin{equation}
\int\limits_{\pa} d\s \, \Sigma_p(\s,\ell|\s_0) < 1  \qquad (\ell > 0),
\end{equation}
i.e., this density is not normalized to $1$, in the same way as the
conventional propagator $G_q(\x,t|\x_0)$ is not normalized to $1$ in
the presence of reactive boundary ($q > 0$).  

In turn, the case $p = 0$ is more subtle.  For restricted diffusion in
a bounded domain, the ground eigenfunction is constant, $v_0^{(0)}(\s)
= |\pa|^{-1/2}$, whereas the associated eigenvalue is zero:
$\mu_0^{(0)} = 0$.  Due to the orthogonality of other eigenfunctions
to $v_0^{(0)}$, the surface hopping propagator is normalized to $1$:
\begin{equation}
\int\limits_{\pa} d\s \, \Sigma_0(\s,\ell|\s_0) = 1  \qquad (\ell > 0).
\end{equation}
In contrast, if diffusion is transient, all eigenvalues $\mu_n^{(0)}$
are strictly positive, and the normalization is lost again, here, due
to the possibility of escaping at infinity.  This is the case of
diffusion in the exterior of a bounded domain in $\R^d$ with $d \geq
3$ (for $d = 2$, see a short discussion in Sec. \ref{sec:diskE}).

\subsection{Relation to other quantities}
\label{sec:other}

As shown in \cite{Grebenkov20}, the surface hopping propagator opens
the door to access most common diffusion-reaction characteristics such
as the full propagator, the conventional propagator, the first-passage
time distribution, and the reaction rate.  In particular, the Laplace
transform of the full propagator reads
\begin{align}  \label{eq:Pfull2}
& \tilde{P}(\x,\ell,p|\x_0) = \tilde{G}_\infty(\x,p|\x_0) \, \delta(\ell) \\  \nonumber
& + \int\limits_{\pa} d\s_0
\int\limits_{\pa} d\s \, \tilde{j}_\infty(\s,p|\x) \, \frac{\Sigma_p(\s,\ell|\s_0)}{D} \, \tilde{j}_\infty(\s_0,p|\x_0) ,
\end{align}
where $G_\infty(\x,t|\x_0)$ is the propagator for perfectly reactive
boundary (with Dirichlet boundary condition
$G_\infty(\x,t|\x_0)_{|\pa} = 0$),
\begin{equation}
j_\infty(\s,t|\x_0) = - D\biggl(\partial_{\n} G_\infty(\x,t|\x_0)\biggr)_{\x=\s}
\end{equation}
is the probability flux density on that boundary, and tilde denotes
the Laplace transform with respect to time $t$, e.g.,
\begin{equation}
\tilde{P}(\x,\ell,p|\x_0) = \int\limits_0^\infty dt \, e^{-pt} \, P(\x,\ell,t|\x_0).
\end{equation}
Substituting the spectral expansion (\ref{eq:Sigma_M}) into
Eq. (\ref{eq:Pfull2}), one also gets
\begin{align}  \label{eq:Pfull30}
\tilde{P}(\x,\ell,p|\x_0) & = \tilde{G}_\infty(\x,p|\x_0) \, \delta(\ell) \\  \nonumber
& + \frac{1}{D} \sum\limits_n [V_n^{(p)}(\x_0)]^* \, V_n^{(p)}(\x) \, e^{-\mu_n^{(p)} \ell} ,
\end{align}
with
\begin{equation}  \label{eq:Vnp}
V_n^{(p)}(\x) = \int\limits_{\pa} d\s \, \tilde{j}_\infty(\s,p|\x) \, v_n^{(p)}(\s).
\end{equation}

In turn, the full propagator determines most common quantities of
diffusion-influenced reactions, in particular, the conventional
propagator via Eq. (\ref{eq:Gq_P}).  Moreover, one gets the marginal
probability density of the boundary local time $\ell_t$ (see also
\cite{Grebenkov07a,Grebenkov19c}):
\begin{equation}
\rho(\ell,t|\x_0) = \int\limits_\Omega d\x \, P(\x,\ell,t|\x_0),
\end{equation}
and the probability density of the first-crossing time $T_\ell =
\inf\{ t>0 ~:~ \ell_t > \ell\}$ of a level $\ell$ by $\ell_t$:
\begin{equation}  \label{eq:U2}
U(\ell,t|\x_0) = D \int\limits_{\pa} d\s \, P(\s,\ell,t|\x_0).
\end{equation}
The latter determines the probability density of the conventional
first-passage time to a partial reactive boundary as:
\begin{equation}   \label{eq:Hpsi}
H_q(t|\x_0) = \int\limits_0^\infty d\ell \, q\, e^{-q\ell} \, U(\ell,t|\x_0).
\end{equation}
In the Laplace domain, one can use the spectral expansion
(\ref{eq:Pfull30}) to write
\begin{equation}  \label{eq:U2p}
\tilde{U}(\ell,p|\x_0) = \sum\limits_n [V_n^{(p)}(\x_0)]^* \, e^{-\mu_n^{(p)}\ell} \int\limits_{\pa} d\s \, v_n^{(p)}(\s).
\end{equation}
Inverting the Laplace transform, one gets $U(\ell,t|\x_0)$ and thus
gains access via Eq. (\ref{eq:Hpsi}) to the whole family of
probability densities $H_q(t|\x_0)$.

The surface hopping propagator also determines the spread harmonic
measure density, $\omega_q(\s|\x_0)$, which characterizes the boundary
point on a partially reactive boundary, at which reaction occurs
\cite{Grebenkov06,Grebenkov06b,Grebenkov15}.  For a particle started
from $\x_0$, one has
\begin{equation}
\omega_q(\s|\x_0) = \int\limits_{\pa} d\s_0 \, \omega_q(\s|\s_0) \,   \tilde{j}_{\infty}(\s_0,0|\x_0),
\end{equation}
where
\begin{equation}
\omega_q(\s|\s_0) = \int\limits_0^\infty d\ell \, q \, e^{-q\ell} \, \Sigma_0(\s,\ell|\s_0) .
\end{equation}
More generally, the Laplace transform of $q \Sigma_p(\s,\ell|\s_0)$
with respect to the boundary local time $\ell$ yields the probability
density of the reaction point $\s$ on the boundary in the presence of
bulk reactions with the rate $p$.

\subsection{Extension}
\label{sec:extensions}

In the above construction of the surface hopping propagator
$\Sigma_p(\s,\ell|\s_0)$ and the full propagator $P(\x,\ell,t|\x_0)$,
the boundary local time $\ell$ is counted on the whole boundary $\pa$
of the confining domain $\Omega$.  This boundary local time is then
used to incorporate surface reactions, like in Eq. (\ref{eq:Gq_P}).
In certain applications, however, only a subset of the boundary,
$\Gamma \subset \pa$, is reactive, whereas the remaining part
$\pa\backslash \Gamma$ is inert and just passively confines the
diffusing particle inside the domain.  This is a typical case of an
escape through a hole $\Gamma$, or of a target $\Gamma$ surrounded by
a reflecting surface.  As encounters of the particle with the passive
part of the boundary do not matter, one needs to count the boundary
local time only on the reactive part $\Gamma$.

An extension to this setting is straightforward.  In fact, one can
re-define $\ell_t$ through the residence time in a close vicinity of
the reactive part: $\Gamma_a = \{ \x\in\Omega ~:~ |\x - \Gamma| <
a\}$:
\begin{equation}
\ell_t = \lim\limits_{a\to 0} \frac{D}{a} \int\limits_0^t dt' \, \I_{\Gamma_a}(\X_{t'}) .
\end{equation}
The associated surface hopping propagator can be constructed as
earlier by modifying the definition of the Dirichlet-to-Neumann
operator.  In fact, as the passive part of the boundary,
$\pa\backslash \Gamma$, is irrelevant, one can define the
Dirichlet-to-Neumann operator $\M_p^\Gamma$, acting on functions on
$\Gamma$ as $\M_p^\Gamma ~:~ f \to g = (\partial_{\n} w)_{|\Gamma}$,
where $w$ is the solution of the mixed Dirichlet-Neumann boundary
value problem:
\begin{equation}  \label{eq:u_def2}
(p - D\Delta) w = 0 \quad \textrm{in}  ~\Omega, \qquad 
\left\{ \begin{array}{l} w_{|\Gamma} = f, \\  (\partial_{\n} w)_{|\pa\backslash \Gamma} = 0. \\ \end{array} \right.
\end{equation}
Here, the Neumann boundary condition on $\pa\backslash \Gamma$
implements explicitly the reflecting character of the passive part of
the boundary.
In the following, we will present several examples of the surface
hopping propagator for a subset of the boundary.

\section{Explicit results for several confining domains}
\label{sec:examples}

In this section, we illustrate the properties of the surface hopping
propagator and related quantities for several confining domains, for
which the eigenbasis of the Dirichlet-to-Neumann operator can be
explicitly derived.  Even though these spectral properties are known
to experts, we will provide some clarifications to guide readers.  We
start with the half-space, for which all quantities, including the
full propagator, will be derived in closed explicit forms.  Then we
consider two-dimensional circular annuli between two concentric
circles that include as limiting cases the interior and the exterior
of a disk.  An extension of these results to three-dimensional
circular cylinders is briefly presented.  Similarly, we discuss
spherical shells between two concentric spheres that include the
interior and the exterior of a ball.  A numerical computation of
the eigenmodes of the Dirichlet-to-Neumann operator in non-concentric
perforated spherical domains is discussed in
\cite{Grebenkov19d,Grebenkov20c}.  Technical derivations are moved to
Appendices.

\subsection{Half-space}
\label{sec:half}

It is instructive to start with the case of the half-space $\Omega =
\{ \x = (x_1,\ldots,x_d)\in \R^d~:~ x_d > 0\}$, for which all
quantities of interest can be obtained in a closed analytic form.
Even though the boundary $\pa$ of the half-space is not bounded, the
derived formulas can be adapted.  In this case, the spectrum of the
Dirichlet-to-Neumann operator is continuous, and sums over eigenmodes
should be replaced by integrals.  Moreover, the solutions of the
eigenvalue equation (\ref{eq:M_eigen}), $v_n^{(p)}$, are not
$L_2(\pa)$-normalized and thus cannot be called ``eigenfunctions''.
Nevertheless, we will keep this term for $v_n^{(p)}$, bearing in mind
its limitations.

\begin{figure}
\begin{center}
\includegraphics[width=88mm]{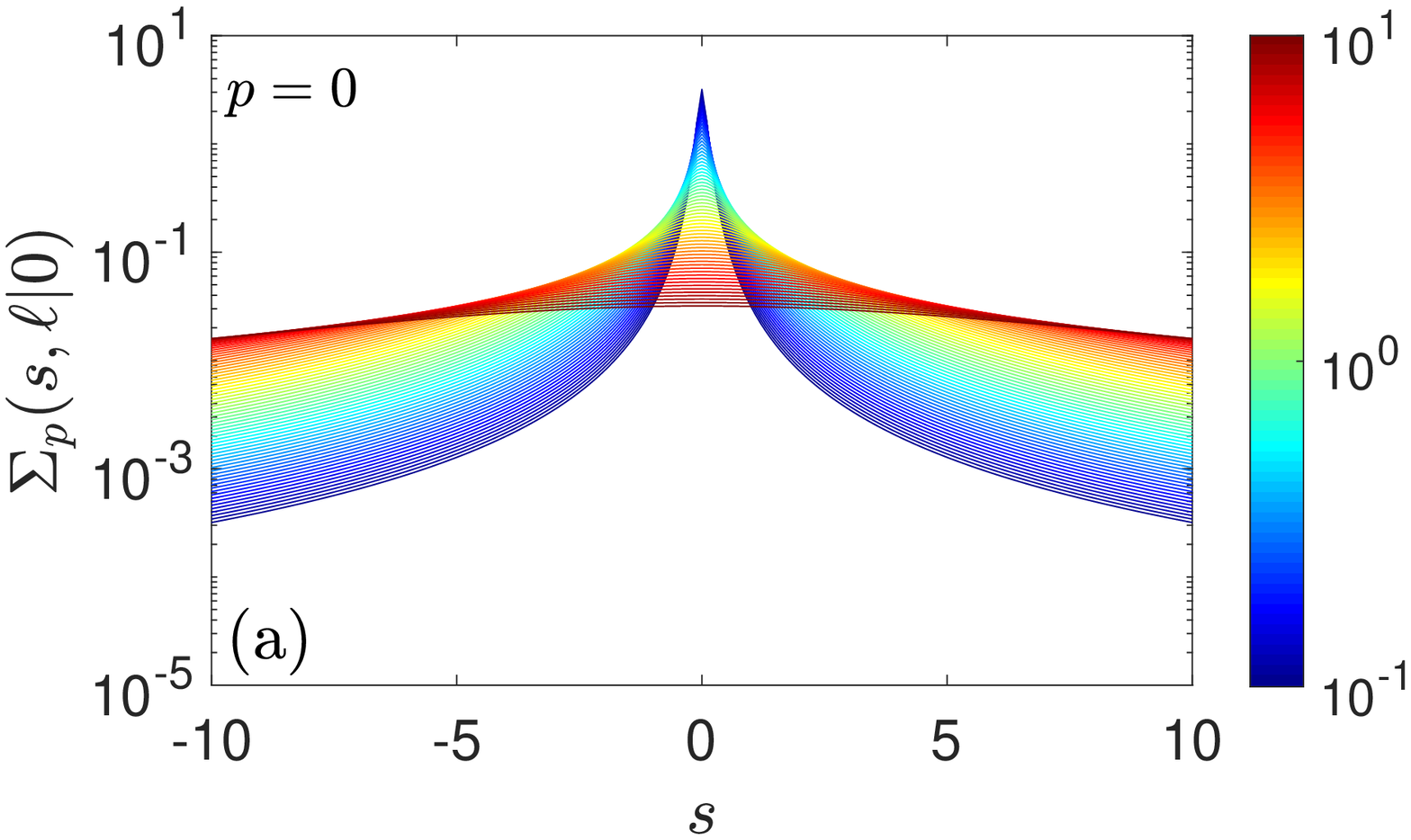} % Sigma_half_p0.eps}
\includegraphics[width=88mm]{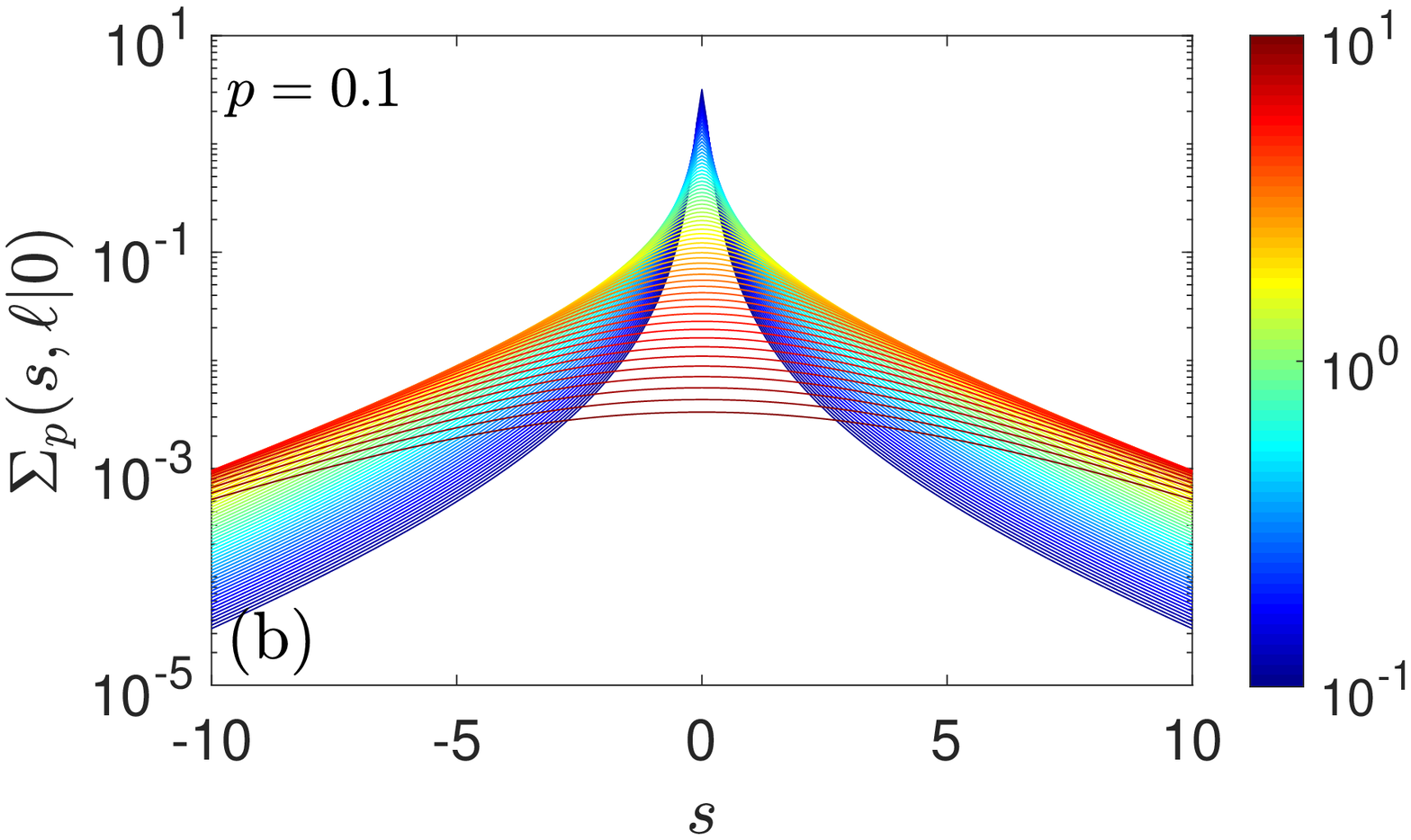} % Sigma_half_p01.eps}
\end{center}
\caption{
The surface hopping propagator $\Sigma_p(s,\ell|0)$, given by
Eq. (\ref{eq:Sigmap_half0}), for diffusion in the upper half-plane for
$p = 0$ {\bf (a)} and $p = 0.1$ {\bf (b)}, and 64 values of $\ell$,
logarithmically spaced in the range from $10^{-1}$ (blue curves) to
$10^1$ (red curves), with $D = 1$.  }
\label{fig:Sigma_half}
% [Sigma, ell,s] = A_localtime_half_Sigma_fig2(0);
% Exported as 20 x 16
\end{figure}

To clarify the ideas, we start again with the planar case ($d = 2$).
The translational symmetry of the boundary implies that $v_n(s) =
e^{ins}/\sqrt{2\pi}$ are the ``eigenfunctions'' of the
Dirichlet-to-Neumann operator $\M_p$.  In fact, since $w(x,y) = e^{inx
- y \sqrt{n^2 + p/D}}$ satisfies the modified Helmholtz equation
(\ref{eq:u_def}) in the upper half-plane, $e^{ins}$ is an
``eigenfunction'', associated to the ``eigenvalue'' $\mu_n^{(p)} =
\sqrt{n^2 + p/D}$.  Note that this ``eigenfunction'' does not depend
on $p$ due to the above symmetry.  The prefactor $1/\sqrt{2\pi}$ comes
from the orthogonality of ``eigenfunctions'':
\begin{equation}
\int\limits_{-\infty}^\infty ds \, v_n(s) \, [v_{n'}(s)]^* = \delta(n-n').
\end{equation}
Skipping technical details, we formally rewrite the spectral expansion
(\ref{eq:Sigma_M}) of the surface hopping propagator as
\begin{equation}
\Sigma_p(s,\ell|s_0) = \int\limits_{-\infty}^\infty dn \, [v_n(s_0)]^* \, v_n(s) \, \exp(-\mu_n^{(p)} \ell) ,
\end{equation}
where the former summation index $n$ now takes real values in $\R$.
As a consequence, we get
\begin{align}  \nonumber
\Sigma_p(s,\ell|s_0) & = \int\limits_{-\infty}^\infty \frac{dn}{2\pi} \, e^{in(s_0 - s) -  \ell \sqrt{n^2 + p/D}} \\  \label{eq:Sigmap_half0}
& = \frac{\ell}{\pi (\ell^2 + (s-s_0)^2)} \, \zeta K_1(\zeta) ,
\end{align}
with 
\begin{equation}  \label{eq:zeta}
\zeta = \sqrt{p/D}\,  \sqrt{\ell^2 + (s-s_0)^2} \,,
\end{equation}
and $K_\nu(z)$ is the modified Bessel function of the second kind.  At
$p = 0$, we retrieve the Cauchy density (\ref{eq:Cauchy}).  In this
case, the surface exploration up to the boundary local time $\ell$ is
equivalent to the first arrival onto that surface of Brownian motion
started from $(s_0,\ell)$, i.e. the distance $\ell$ above the surface.
Figure \ref{fig:Sigma_half} illustrates the behavior of the surface
hopping propagator.  Changing progressively the boundary local time
$\ell$, one observes the spreading of the surface hopping propagator.

For ``immortal'' particles ($p = 0$), the surface hopping propagator
exhibits heavy tails, $\Sigma_0(s,\ell|s_0) \propto |s-s_0|^{-2}$, in
particular, the variance of the arrival point $s$ is infinite (the
mean is $s_0$ due to the symmetric form of this propagator).  Such
displacements with infinite variance resemble L\'evy flights
\cite{Zaburdaev15}.  This is the consequence of unbounded exploration
region that allows for very long and far-reaching trajectories.  The
situation is drastically different for ``mortal'' particles ($p > 0$),
for which long trajectories are penalized by tiny chances of survival.
In fact, the central part of this distribution (when $\ell^2 +
|s-s_0|^2 \ll D/p$) resembles again the Cauchy density (Fig.
\ref{fig:Sigma_half}(b)), which is however truncated by exponential
tails at large $s$ (when $\zeta \gg 1$): $\Sigma_p(s,\ell|s_0) \propto
e^{-\zeta} \sim e^{-|s-s_0|\sqrt{p/D}}$.  The bulk rate $p$ (or, more
precisely, $p/D$) controls this truncation.  As a consequence, all the
positive moments of the arrival boundary point are finite.
Expectedly, the surface hopping propagator is not normalized to $1$
for $p > 0$:
\begin{equation}
\int\limits_{\R} ds \, \Sigma_p(s,\ell|s_0) = e^{-\ell \sqrt{p/D}} \,.
\end{equation}
Interestingly, even the conditional surface hopping propagator that
accounts only for the survived particles after renormalization by
$e^{-\ell\sqrt{p/D}}$, shows an exponential decay with $s$.

In higher dimensions, the ``eigenfunctions'' and ``eigenvalues'' of
the Dirichlet-to-Neumann operator have similar form:
\begin{equation}
v_{\n}(\s) = \frac{e^{i(\n\cdot\s)}}{(2\pi)^{(d-1)/2}} \,, \qquad 
\mu_{\n}^{(p)} = \sqrt{|\n|^2 + p/D}, 
\end{equation}
with the ``multi-index'' $\n = (n_1,\ldots,n_{d-1}) \in
\R^{d-1}$.  The surface propagator can thus be written as
\begin{equation}  \label{eq:Sigma_half1}
\Sigma_p(\s,\ell|\s_0) = \int\limits_{\R^{d-1}} \frac{d\n}{(2\pi)^{d-1}} \, e^{i\n\cdot (\s_0 -\s) - \ell \sqrt{|\n|^2 + p/D}} \,.
\end{equation}
In spherical coordinates, the integral over all orientations gives
\begin{align}  \nonumber
\Sigma_p(\s,\ell|\s_0) & = \frac{|\s-\s_0|^{\frac{3-d}{2}}}{(2\pi)^{\frac{d-1}{2}}} 
\int\limits_0^\infty dn \, n^{\frac{d-1}{2}} J_{\frac{d-3}{2}}(n|\s-\s_0|) \\  \nonumber
& \times e^{-\ell\sqrt{n^2 + p/D}} \\  \label{eq:Sigmap_half}
& = \Sigma_0(\s,\ell|\s_0) \, \frac{\zeta^{\frac{d}{2}} \, K_{\frac{d}{2}}(\zeta)}{\Gamma(\frac{d}{2}) 2^{\frac{d}{2}-1}} \,,
\end{align}  % checked in [S, S0, p, zeta] = A_localtime_check2(d);
where $\zeta$ is given by Eq. (\ref{eq:zeta}), and
\begin{equation}
\Sigma_0(\s,\ell|\s_0) = \frac{\Gamma(d/2)}{\pi^{d/2}} \, \frac{\ell}{(\ell^2 + |\s-\s_0|^2)^{d/2}}  
\end{equation}
is again the harmonic measure density on the hyperplane $\R^{d-1}$
(see also \cite{Grebenkov15}), and $J_\nu(z)$ is the Bessel function
of the first kind.  In the right-hand side of
Eq. (\ref{eq:Sigmap_half}), one can recognize the Laplace-transformed
probability flux density $ \tilde{j}_\infty(\s,p|(\s_0,\ell))$ onto a
perfectly absorbing hyperplane from the bulk point $\x_0 = (\s_0,\ell)$.
The inverse Laplace transform with respect to $p$ yields then
\begin{align}  \label{eq:jinf_half}
\L^{-1}_t\{ \Sigma_p(\s,\ell|\s_0)\} & = j_\infty(\s,t|(\s_0,\ell)) \\   \nonumber
& = \frac{\exp\bigl(-\frac{|\s - \s_0|^2}{4Dt}\bigr)}{(4\pi D t)^{(d-1)/2}} \, 
\frac{\ell \, \exp\bigl(-\frac{\ell^2}{4Dt}\bigr)}{\sqrt{4\pi Dt^3}} .
\end{align}
We emphasize that this relation is specific to the case of the
half-space.

Using this relation and the representation (\ref{eq:Sigma_half1}), one
can easily compute the double integral over $\s_1$ and $\s_2$ in
Eq. (\ref{eq:Pfull30}) to get
\begin{equation}
\tilde{P}(\x,\ell,p|\x_0) = \tilde{G}_\infty(\x,p|\x_0) \delta(\ell) + \frac{\Sigma_p(\y,z+z_0+\ell|\y_0)}{D} ,
\end{equation}
where $\x = (\y,z)$ and $\x_0 = (\y_0,z_0)$.  The inverse Laplace
transform with respect to $p$ yields then
\begin{align}  \nonumber
& P(\x,\ell,t|\x_0) = G_\infty(\x,t|\x_0) \delta(\ell) + \frac{j_\infty(\y,t|(\y_0,z+z_0+\ell))}{D} \\  \nonumber
& = \frac{\exp\bigl(-\frac{|\y-\y_0|^2}{4Dt}\bigr)}{(4\pi Dt)^{d/2}} 
\biggl\{ \bigl( e^{-(z-z_0)^2/(4Dt)} - e^{-(z+z_0)^2/(4Dt)} \bigr) \delta(\ell) \\   \label{eq:P_half}
& + \frac{\ell + z + z_0}{Dt} e^{-(z+z_0+\ell)^2/(4Dt)} \biggr\}.
\end{align}
This is the explicit exact form of the full propagator for the
half-space.

One can easily check that the integral over $\x \in \R^d_+$ gives the
marginal probability density of the boundary local time $\ell_t$:
\begin{equation}
\rho(\ell,t|\x_0) = \erf\biggl(\frac{z_0}{\sqrt{4Dt}}\biggr) \delta(\ell) + \frac{\exp\bigl(- \frac{(z_0+\ell)^2}{4Dt}\bigr)}{\sqrt{\pi D t}} \,.
\end{equation}
This expression does not depend on the dimension $d$ and the lateral
coordinate $\y_0$ of the starting point $\x_0$, given that the
boundary local time is independent of lateral displacements and
determined by the transverse motion (on the half-line).  The
distribution of the boundary local time $\ell_t$ was studied in
\cite{Grebenkov07a,Grebenkov19c}.  

In turn, the integral of Eq. (\ref{eq:P_half}) over $\ell \in \R_+$
yields the marginal probability density of the position, i.e.,
conventional propagator $G_0(\x,t|\x_0)$ in the half-space with
reflecting boundary:
\begin{align}  \nonumber
G_0(\x,t|\x_0) & = \frac{\exp\bigl(-\frac{|\y-\y_0|^2}{4Dt}\bigr)}{(4\pi Dt)^{d/2}} \\  \label{eq:Gq_half0}
& \times \bigl( e^{-(z-z_0)^2/(4Dt)} + e^{-(z+z_0)^2/(4Dt)} \bigr).
\end{align}
Moreover, with the general expression (\ref{eq:Gq_P}), one retrieves
the propagator $G_q(\x,t|\x_0)$ for reactive boundary:
\begin{align}  \nonumber
& G_q(\x,t|\x_0) = \frac{\exp\bigl(-\frac{|\y-\y_0|^2}{4Dt}\bigr)}{(4\pi Dt)^{d/2}} 
\biggl\{ e^{-(z-z_0)^2/(4Dt)} + \\  \label{eq:Gq_half}
& e^{-(z+z_0)^2/(4Dt)} \biggl(1 - 2q\sqrt{\pi Dt}\, \erfcx\biggl(\frac{z+z_0}{\sqrt{4Dt}} + q\sqrt{Dt}\biggr)\biggr)\biggr\},
\end{align}
where $\erfcx(z) = e^{z^2} \erfc(z)$ is the scaled complementary error
function.  Expectedly, the propagators in Eqs. (\ref{eq:Gq_half0},
\ref{eq:Gq_half}) exhibit translational invariance along $\y$
coordinate and are factored into the lateral Gaussian (free)
propagator and the transverse propagator on the half-line
$(0,\infty)$.  Note that its integral over the arrival point $\x$
yields the survival probability
\begin{equation}
S_q(t|\x_0) = \erf\biggl(\frac{z_0}{\sqrt{4Dt}}\biggr) + 
e^{-\frac{z_0^2}{4Dt}} \erfcx\biggl(\frac{z_0}{\sqrt{4Dt}} + q\sqrt{Dt}\biggr),
\end{equation}
which does not depend on $\y_0$ and coincides with the survival 
probability for the semi-axis with the partially reactive endpoint.
The classical expression for the associated probability density of 
the reaction time is then retrieved:
\begin{align}  \nonumber
H_q(t|\x_0) & = qD e^{-z_0^2/(4Dt)} \biggl\{\frac{1}{\sqrt{\pi Dt}} \\
& - q\, \erfcx\biggl(\frac{z_0}{\sqrt{4Dt}} + q\sqrt{Dt}\biggr) \biggr\} .
\end{align}
Finally, Eq. (\ref{eq:U2}) yields
\begin{equation}  
U(\ell,t|\x_0) = \frac{(\ell+z_0) e^{-(\ell+z_0)^2/(4Dt)}}{\sqrt{4\pi D t^3}}  \,,
\end{equation}
i.e., we retrieved the classical formula for the probability density
of the first crossing time of a level $\ell$ by the boundary local
time of reflected Brownian motion on the half-line (see, e.g.,
\cite{Borodin}).

To complete this section, we briefly mention that the above
computations can be easily extended to a slab domain between parallel
hyperplanes, one of which is reflecting (see \cite{Grebenkov20b} for
more details).  In other words, one can consider $\Omega =
\{\x\in \R^d ~:~ 0 < x_d < L \}$ and study the Dirichlet-to-Neumann
operator on the hyperplane $x_d = 0$ in the presence of the reflecting
hyperplane $x_d = L$.  The ``eigenfunctions'' of such
Dirichlet-to-Neumann operator remain unchanged due to the
translational symmetry, whereas the ``eigenvalues'' are
\begin{equation}  \label{eq:mu_slab}
\mu_{\n}^{(p)} = \sqrt{|\n|^2 + p/D} \, \tanh\bigl(L \sqrt{|\n|^2 + p/D}\bigr).
\end{equation}
As $L\to\infty$, one retrieves the former case of the half-space.  The
former integral representations for the surface hopping propagator and
related quantities remain valid if Eq. (\ref{eq:mu_slab}) is used for
$\mu_{\n}^{(p)}$.  In contrast, the presence of $\tanh(z)$ prevents
from getting simple closed formulas for these quantities in the case
of a slab.

\subsection{Circular annuli and spherical shells}

In Appendices \ref{sec:annulus} and \ref{sec:shell}, we provide
explicit formulas for the eigenfunctions and eigenvalues of the
Dirichlet-to-Neumann operator in several rotationally invariant
domains: circular annuli, the interior and the exterior of a disk,
circular cylinders, spherical shells, the interior and the exterior of
a ball.  These formulas allow one to get the surface hopping
propagator $\Sigma_p(\s,\ell|\s_0)$ via the spectral expansion
(\ref{eq:Sigma_M}), as well as the full propagator $P(\x,\ell,t|\x_0)$
and all the related quantities, as discussed in Sec. \ref{sec:other}.
From these basic results, one can thoroughly investigate various
diffusion-mediated surface phenomena in the above domains.  In this
paper, we keep our focus on the surface hopping propagator and
illustrate its properties for these domains.

\begin{figure}
\begin{center}
\includegraphics[width=42mm]{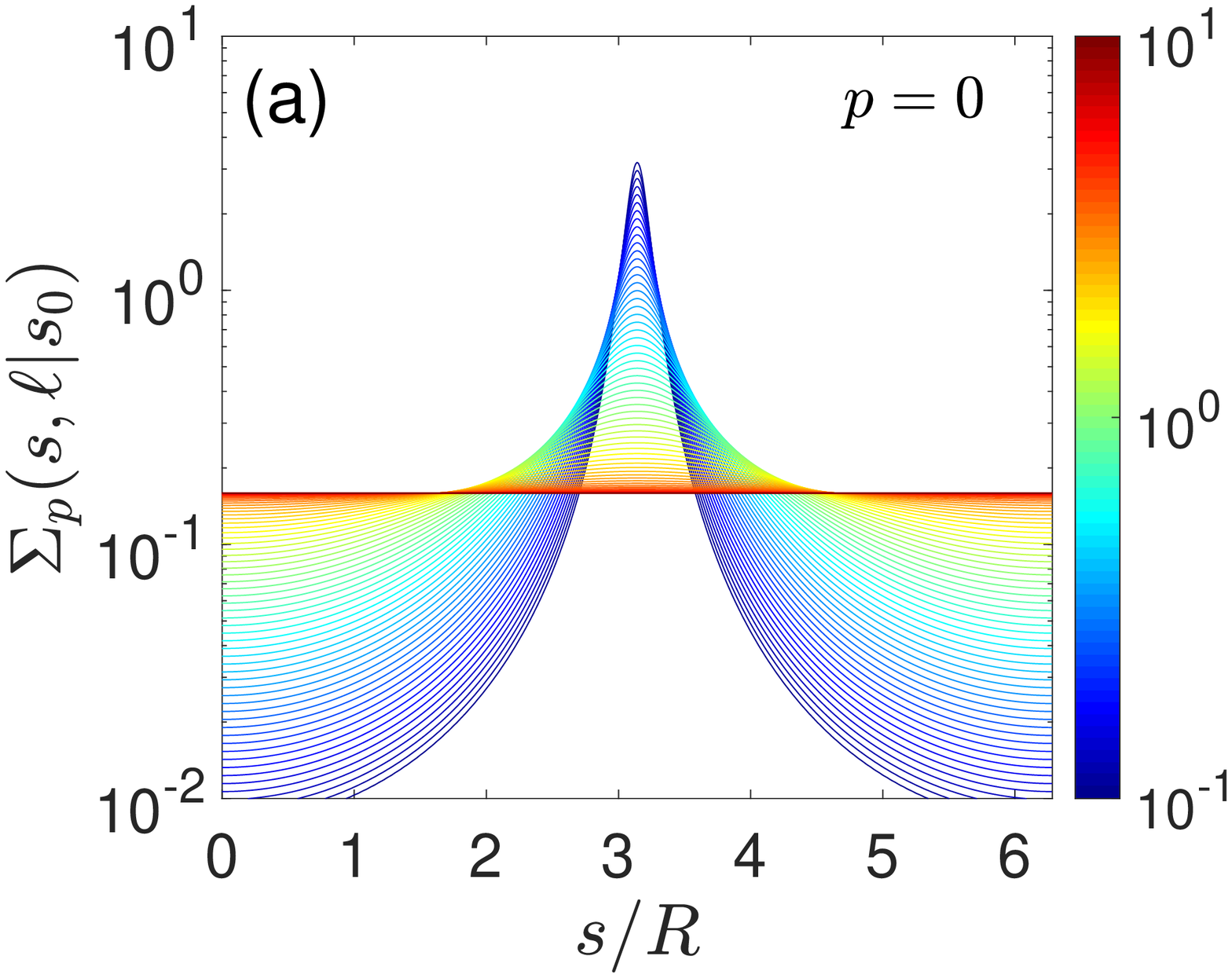} % Sigma_diskI_p0.eps}
\includegraphics[width=42mm]{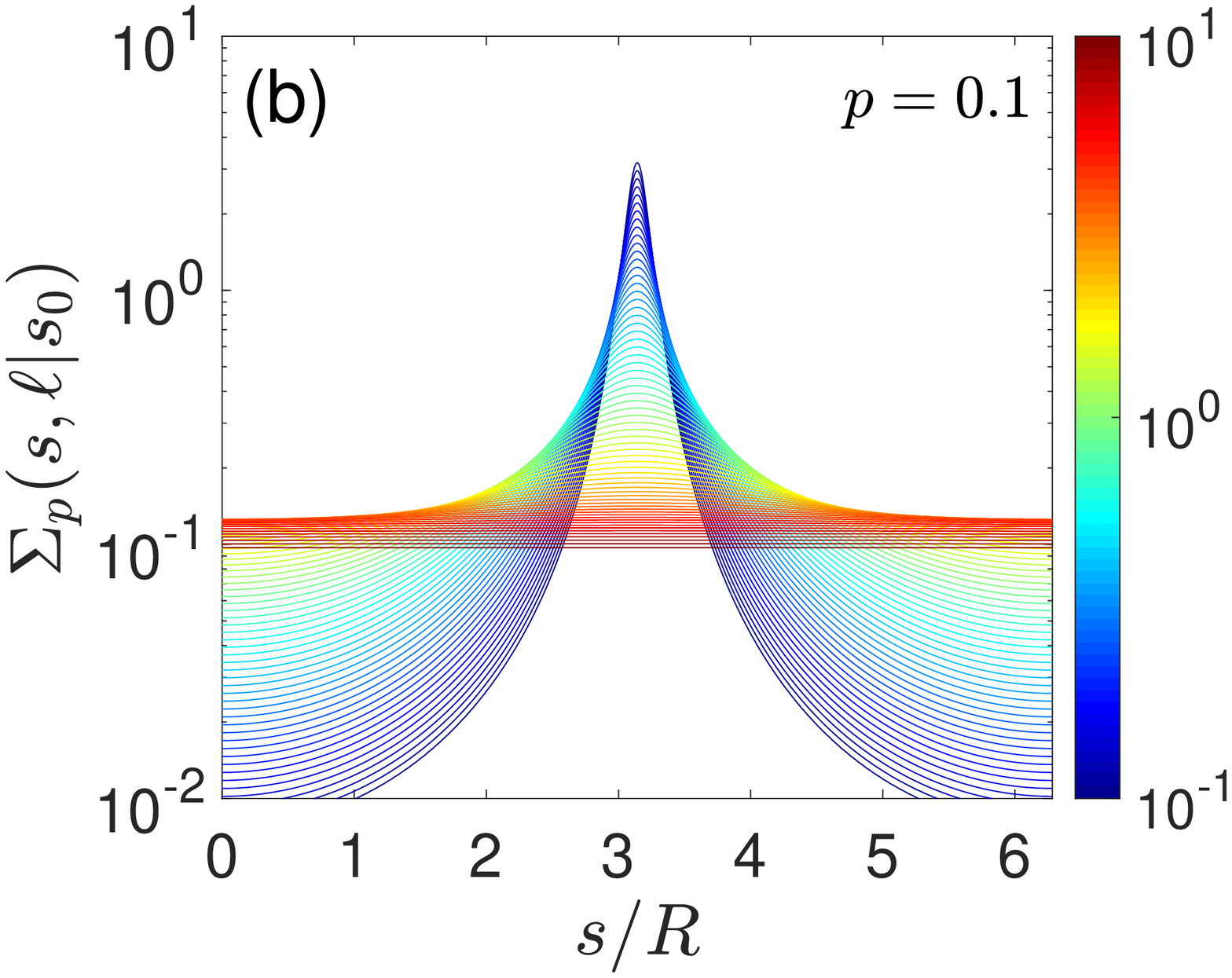} % Sigma_diskI_p01.eps}
\includegraphics[width=42mm]{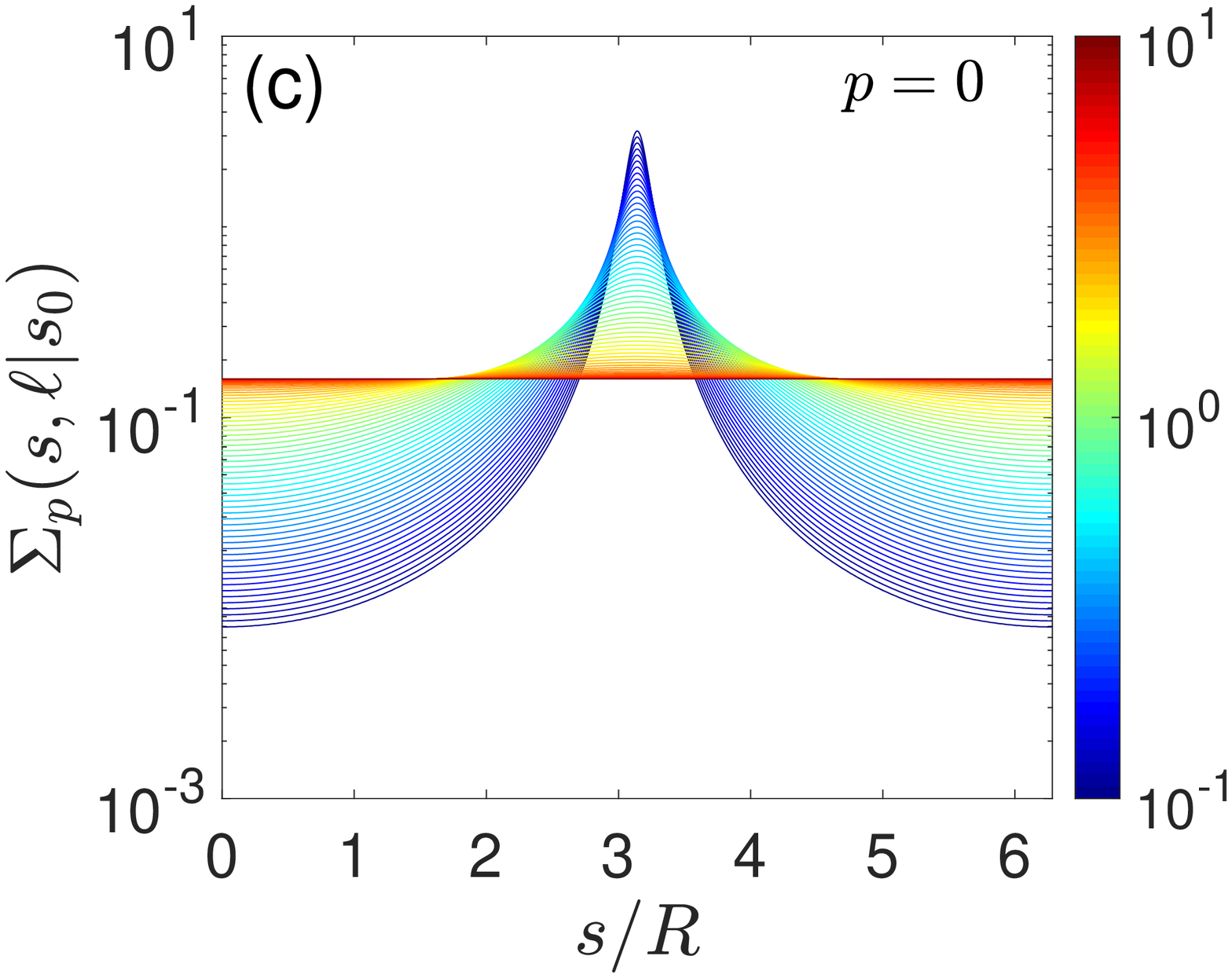} % Sigma_diskE_p0.eps}
\includegraphics[width=42mm]{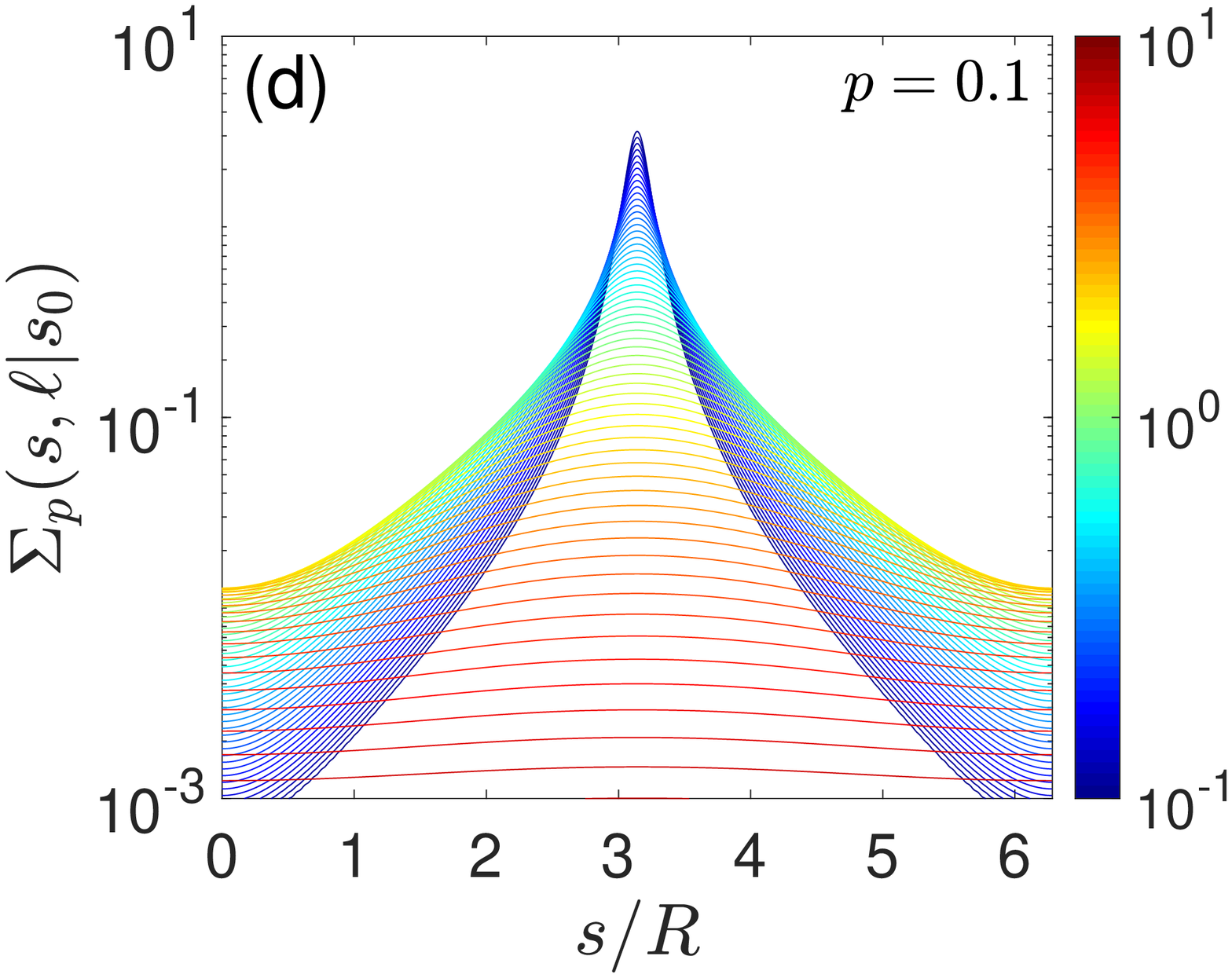} % Sigma_diskE_p01.eps}
\end{center}
\caption{
{\bf (a,b)} The surface hopping propagator $\Sigma_p(s,\ell|s_0)$,
given by Eq. (\ref{eq:Sigmap_annulus}), for diffusion inside a disk of
radius $R$, for $p = 0$ {\bf (a)} and $p = 0.1$ {\bf (b)}, and 64
values of $\ell$, logarithmically spaced in the range from $10^{-1}$
(blue curves) to $10^1$ (red curves), with $D = 1$ and $s_0/R = \pi$.
{\bf (c,d)} The surface hopping propagator $\Sigma_p(s,\ell|s_0)$ for
diffusion outside a disk of radius $R$, with the same parameters.
Note that the propagators on panels {\bf (a)} and {\bf (c)} are
identical but the vertical axis is cut differently.  The series in
Eq. (\ref{eq:Sigmap_annulus}) is truncated above $|n| = 100$.}
\label{fig:Sigma_disk}
% [Sigma, ell,s] = A_localtime_Sigma_fig;
% Exported as 20 x 16
\end{figure}

Figure \ref{fig:Sigma_disk}(a,b) shows the surface hopping propagator
for the interior of a disk of radius $R$.  At $p = 0$ (panel (a)), the
surface hopping propagator $\Sigma_0(s,\ell|s_0)$ coincides with the
harmonic measure density on the circle, see Eq. (\ref{eq:Sigma_disk}).
Expectedly, it evolves from the Dirac distribution at $\ell = 0$ to
the uniform distribution $1/(2\pi R)$ as $\ell\to \infty$.  When $p >
0$ (panel (b)), eventual death of the particle during its bulk
explorations affects this propagator.  At small $\ell$, the particle
spends short time in the bulk so that there is almost no effect of the
bulk rate $p$ (at moderate $p$): blue curves on (a) and (b) panels are
almost identical.  As $\ell$ increases, the effect of $p$ becomes more
prominent.  At large $\ell$, the particle has enough time to explore
the interior of the disk, leading again to the uniform distribution of
the arrival point.  This can also be seen from the spectral expansion
(\ref{eq:Sigma_M}), in which the contribution from higher-order
eigenmodes of the Dirichlet-to-Neumann operator vanished.  The surface
hopping propagator becomes almost flat again, $\Sigma_p(s,\ell|s_0)
\simeq e^{-\mu_0^{(p)}\ell}/(2\pi R)$, but its level is now attenuated
by bulk reactions.  As $\mu_0^{(p=0.1)} \approx 0.0494$, the
attenuation factor $e^{-\mu_0^{(p)}\ell}$ is not strong even at $\ell
= 10$ in the considered example.  However, one can still see a
qualitative difference between $p = 0$ and $p = 0.1$ cases: in the
former case, the curves approach the limit $1/(2\pi R)$, whereas in
the latter case, they are progressively shifted downward.

Figure \ref{fig:Sigma_disk}(c,d) shows the surface hopping propagator
for the exterior of a disk of radius $R$.  We note that
Eq. (\ref{eq:Sigma_disk}) for the surface hopping propagator at $p =
0$ remains the same for diffusion outside the disk.  This reflects the
conformal invariance of the harmonic measure with respect to an
inversion mapping of the interior of the disk to its exterior.  In
contrast, for $p > 0$, the behavior of the surface hopping propagator
is different for diffusion inside and outside the disk, especially for
large $\ell$.  In fact, the particle diffusing outside the disk
undertakes much longer bulk excursions between successive encounters
with the boundary and thus has higher chances to be killed by a bulk
reaction.  This leads to much smaller values of the surface hopping
propagator (we recall that the propagator is not normalized to $1$ for
$p > 0$).

\begin{figure}
\begin{center}
\includegraphics[width=88mm]{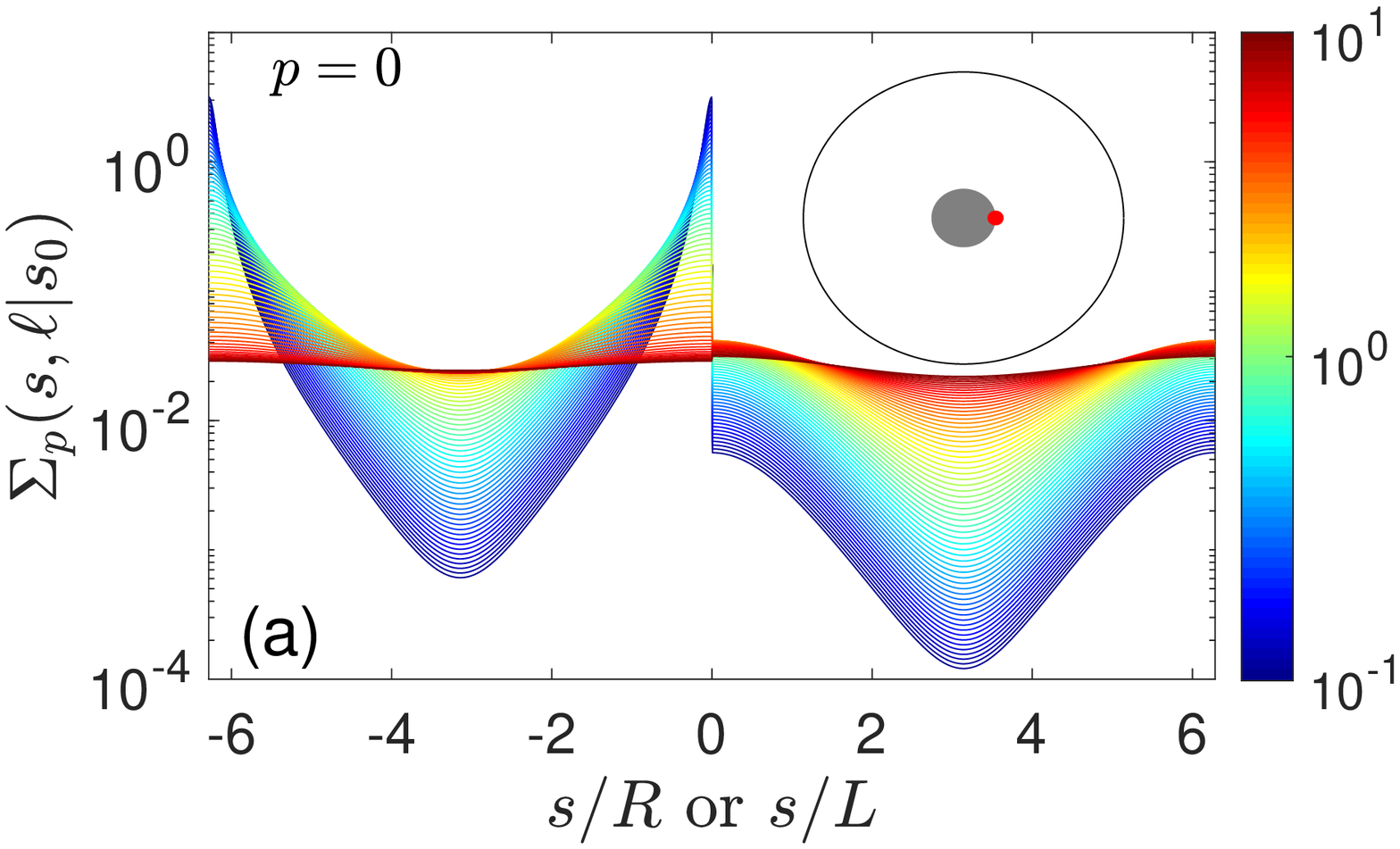} % Sigma_shellDD1_p0.eps}
\includegraphics[width=88mm]{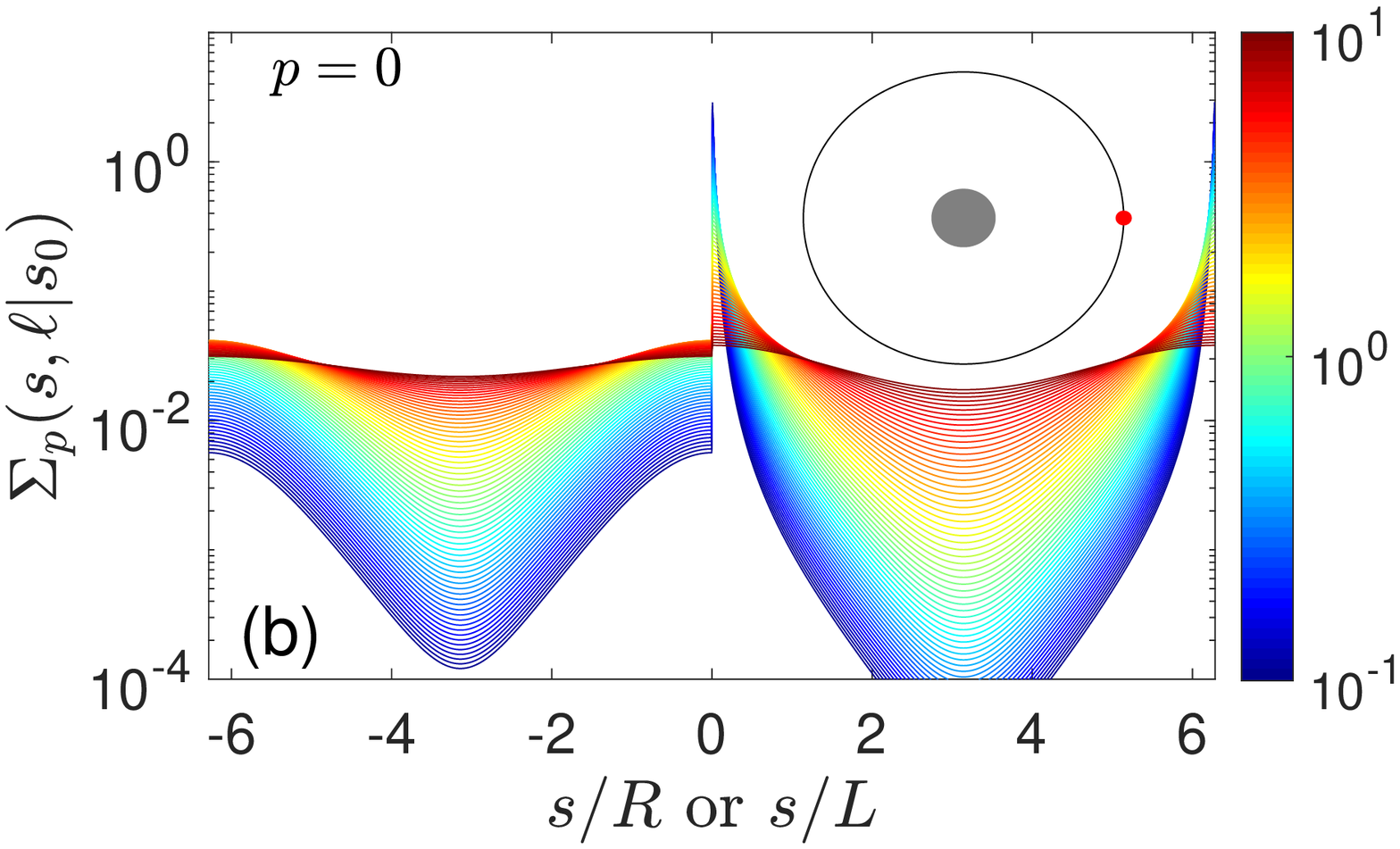} % Sigma_shellDD2_p0.eps}
\end{center}
\caption{
The surface hopping propagator $\Sigma_p(s,\ell|s_0)$ for a circular
annulus with both reactive circles of radii $R$ and $L$, for 64 values
of $\ell$, logarithmically spaced in the range from $10^{-1}$ (blue
curves) to $10^1$ (red curves), with $p = 0$, $L = 5R$, $D = 1$.  {\bf
(a)} The starting point $s_0 = 0$ (red dot) is on the inner circle;
{\bf (b)} the starting point $s_0 = 0$ is on the outer circle.  The
series is truncated above $|n| = 1000$. }
\label{fig:Sigma_shellDD}
% [Sigma1,Sigma2, ell,s] = A_localtime_SigmaDD_fig(Sigma1,Sigma2);
\end{figure}

Figure \ref{fig:Sigma_shellDD} presents the surface hopping propagator
$\Sigma_0(s,\ell|s_0)$ for a circular annulus with both reactive
circles of radii $R$ and $L$.  As the boundary is composed of two
circles, there are two distinct choices of the starting point $s_0$
(shown by red dot): either on the inner circle of radius $R$, or on
the outer circle of radius $L$ (and one can set $s_0 = 0$ in both
cases due to rotational invariance).  For convenience of presentation,
the curvilinear coordinate $s$ runs here from $-2\pi R$ to $0$ for the
inner circle, and from $0$ to $2\pi L$ for the outer circle.  On the
horizontal axis, $s$ is rescaled by $R$ for negative values and by $L$
for positive values so that the horizontal axis varies from $-2\pi$
and $2\pi$.  Let us first consider the particle started on the inner
circle (Fig. \ref{fig:Sigma_shellDD}(a)).  At small $\ell$ (blue
curves), this particle does not move far away from the starting point
$s_0$, so that $\Sigma_0(s,\ell|s_0)$ rapidly decays when $s/R$ varies
from $0$ to $-\pi$ (its later increase for $s/R$ ranging from $-\pi$
to $-2\pi$ is due to the symmetry).  Similarly, bulk excursions of the
particle rarely terminate at the outer circle so that the surface
hopping propagator remains small on that boundary (the range of
positive $s$).  Clearly, the minimum corresponds to the boundary point
$s = \pi L$ on the outer boundary which is located behind the starting
point $s_0$.  As the boundary local time increases, the particle
explores further boundary regions, both on the inner and outer
circles.  In the limit $\ell \to \infty$, the surface hopping
propagator approaches the uniform density, $1/(2\pi(R+L))$, on both
circles, as expected.
When the starting point is on the outer circle
(Fig. \ref{fig:Sigma_shellDD}(b)), the picture is very similar, i.e.,
the particle remains on the outer circle (close to the starting point)
at small $\ell$ but then spreads away.  Note that the stronger decay
of the surface hopping propagator is caused by the fact that the outer
circle is much longer than the inner one.  For this reason, a large
truncation order was needed to accurately compute
$\Sigma_0(s,\ell|s_0)$ in this case.

For ``mortal'' particles (not shown), eventual death due to the bulk
rate $p > 0$ penalizes long trajectories, as in the case of diffusion
inside a disk.  Moreover, in highly reactive media, the particle has
tiny chances to move from one circle to the other, and these parts of
the boundary become decoupled.  In other words, as there is almost no
survived particles that crossed the annulus, the properties of the
boundary far away from the starting point do not matter.

\begin{figure}
\begin{center}
\includegraphics[width=42mm]{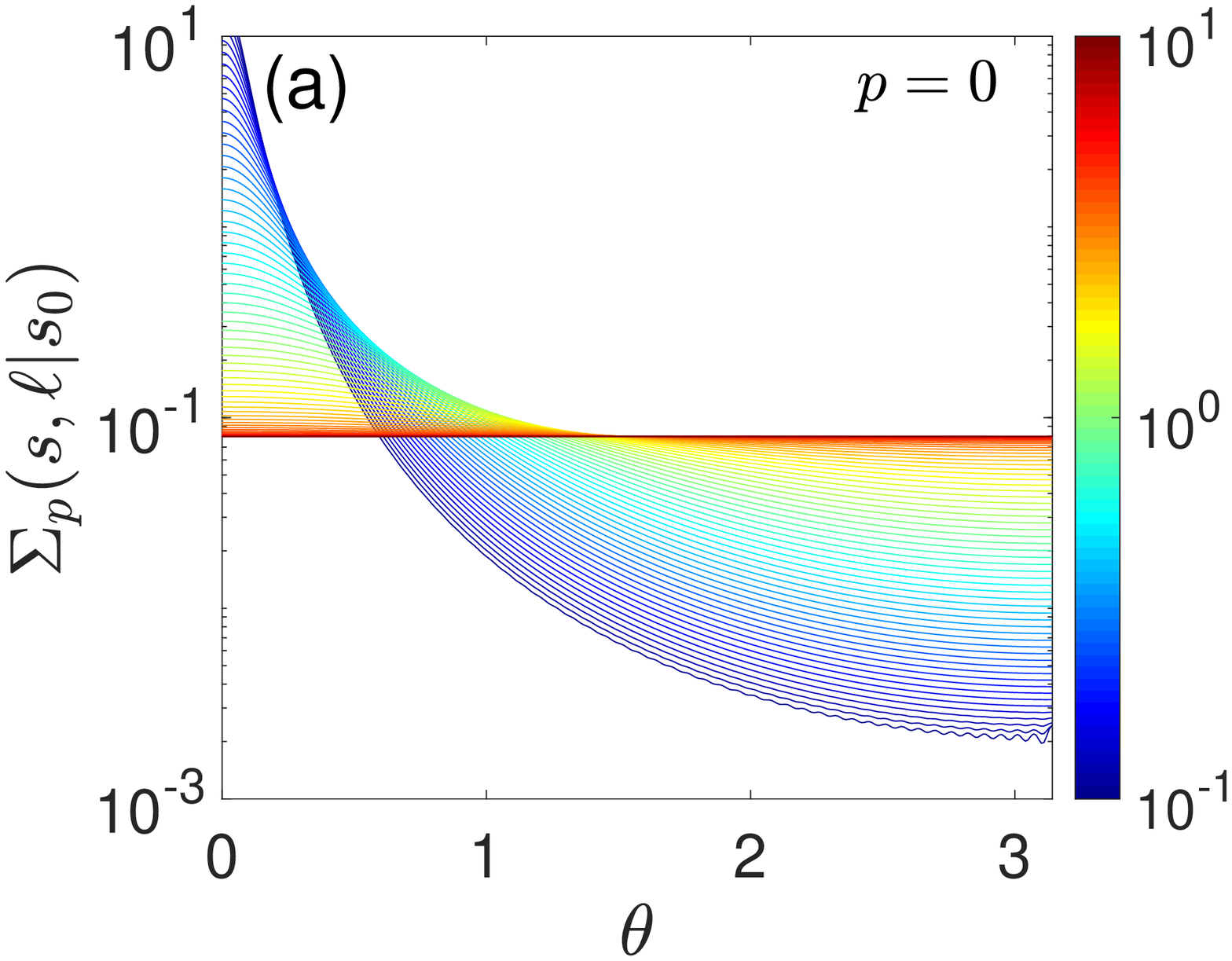} % Sigma_ballI_p0.eps}
\includegraphics[width=42mm]{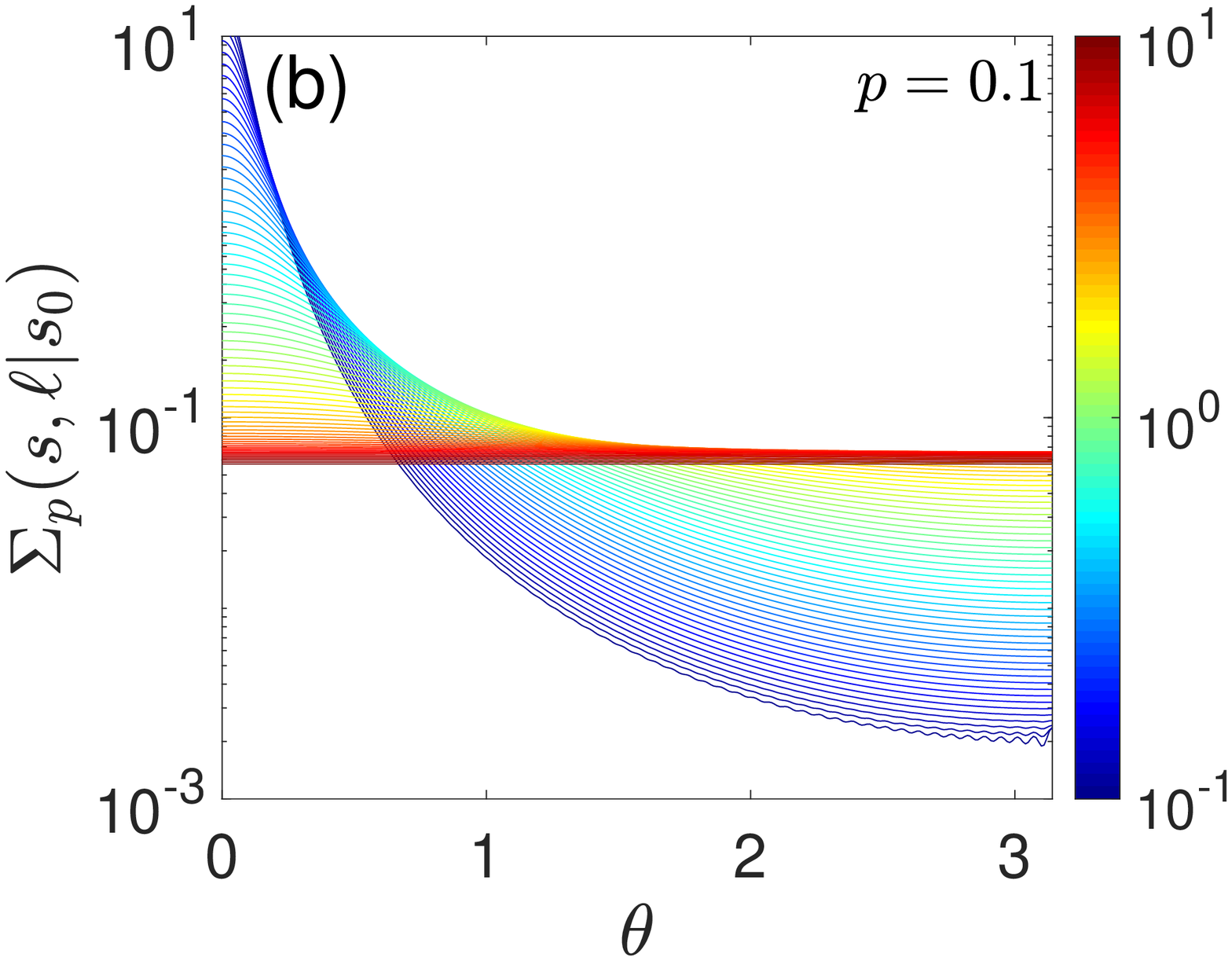} % Sigma_ballI_p01.eps}
\includegraphics[width=42mm]{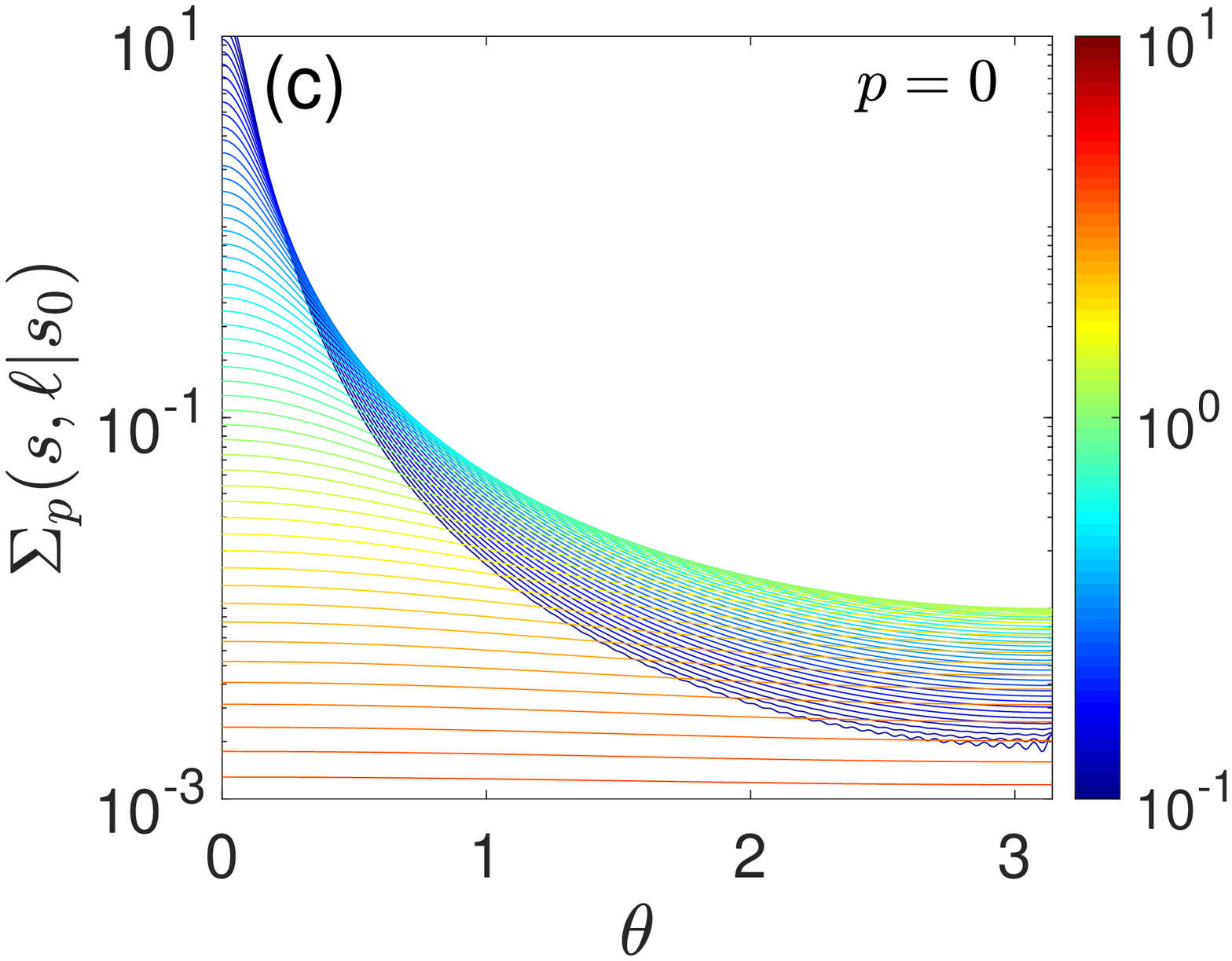} % Sigma_ballE_p0.eps}
\includegraphics[width=42mm]{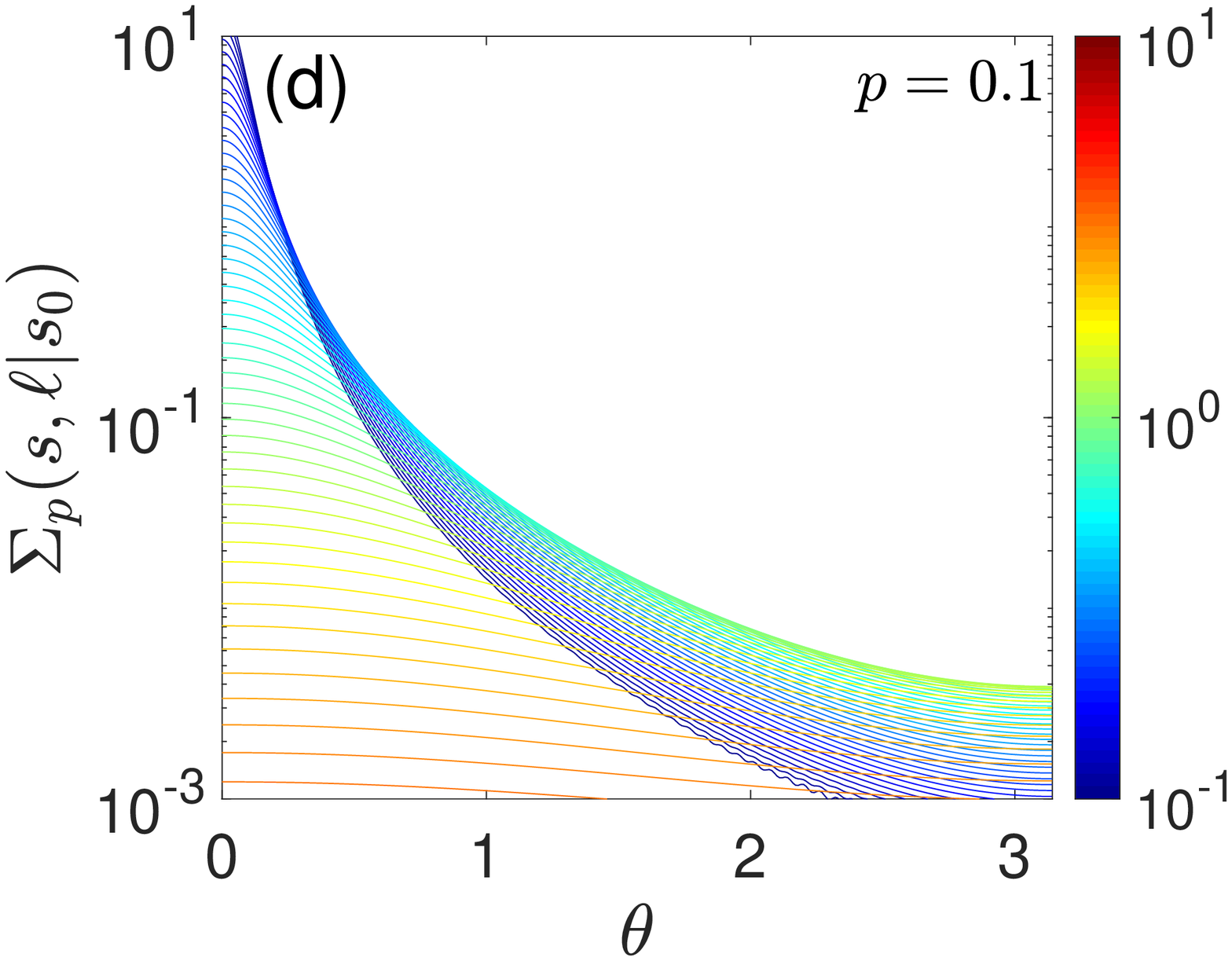} % Sigma_ballE_p01.eps}
\end{center}
\caption{
{\bf (a,b)} The surface hopping propagator $\Sigma_p(\s,\ell|\s_0)$,
given by Eq. (\ref{eq:Sigma_shell2}), for diffusion inside a ball of
radius $R$, for $p = 0$ {\bf (a)} and $p = 0.1$ {\bf (b)}, and 64
values of $\ell$, logarithmically spaced in the range from $10^{-1}$
(blue curves) to $10^1$ (red curves), with $D = 1$, $\s_0 = (0,0,R)$
(the North pole), and $\s = (R\sin \theta,0,R\cos\theta)$ (i.e.,
$\theta$ is the angle between $\s_0$ and $\s$).  {\bf (c,d)} The
surface hopping propagator $\Sigma_p(\s,\ell|\s_0)$ for diffusion
outside a ball of radius $R$, with the same parameters.  The series in
Eq. (\ref{eq:Sigma_shell2}) is truncated above $n = 100$ that results
in small oscillations seen for blue curves. }
\label{fig:Sigma_ball}
% [Sigma, ell,s] = A_localtime_Sigma_fig;
% Exported as 20 x 16
\end{figure}

Figure \ref{fig:Sigma_ball} shows similar results for the interior and
exterior of a ball.  As previously, $\Sigma_p(\s,\ell|\s_0)$ behaves
differently for diffusion inside and outside the ball.  However, this
difference is considerably enhanced in three dimensions due to the
recurrent versus transient character of diffusion.  In fact, the
particle diffusing outside a ball can escape to infinity with a finite
probability.  As a consequence, the surface hopping propagator is not
normalized to $1$ here even for $p = 0$.  Actually, the normalization
constant is $e^{-\ell/R}$, i.e., the escape probability is $1 -
e^{-\ell/R}$.  On panel (c), one can see that the surface hopping
propagator becomes again uniform (as the contribution of higher-order
eigenfunctions vanishes) but attenuated by the factor $e^{-\ell/R}$.

\begin{figure}
\begin{center}
\includegraphics[width=42mm]{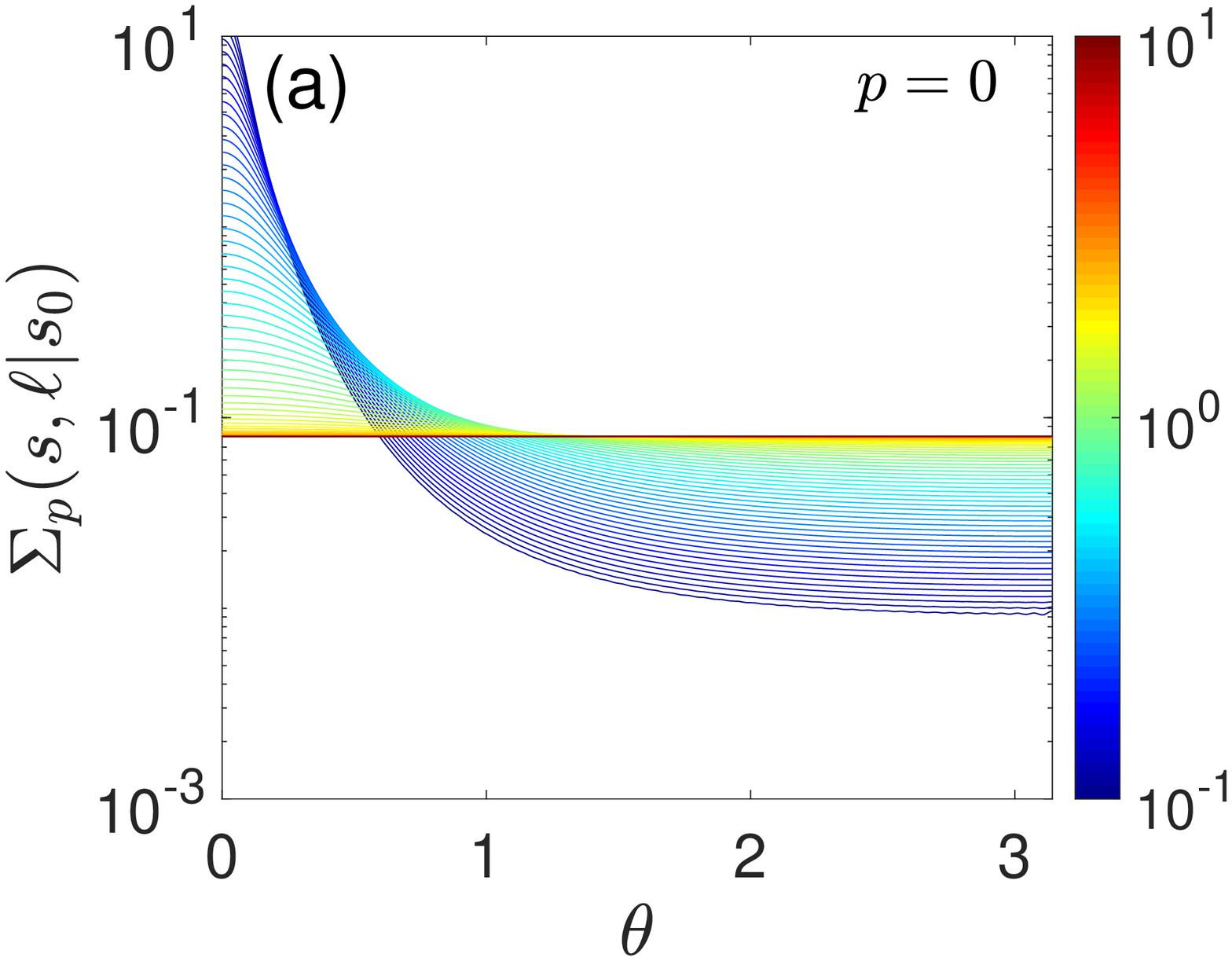} % Sigma_shell_p0.eps}
\includegraphics[width=42mm]{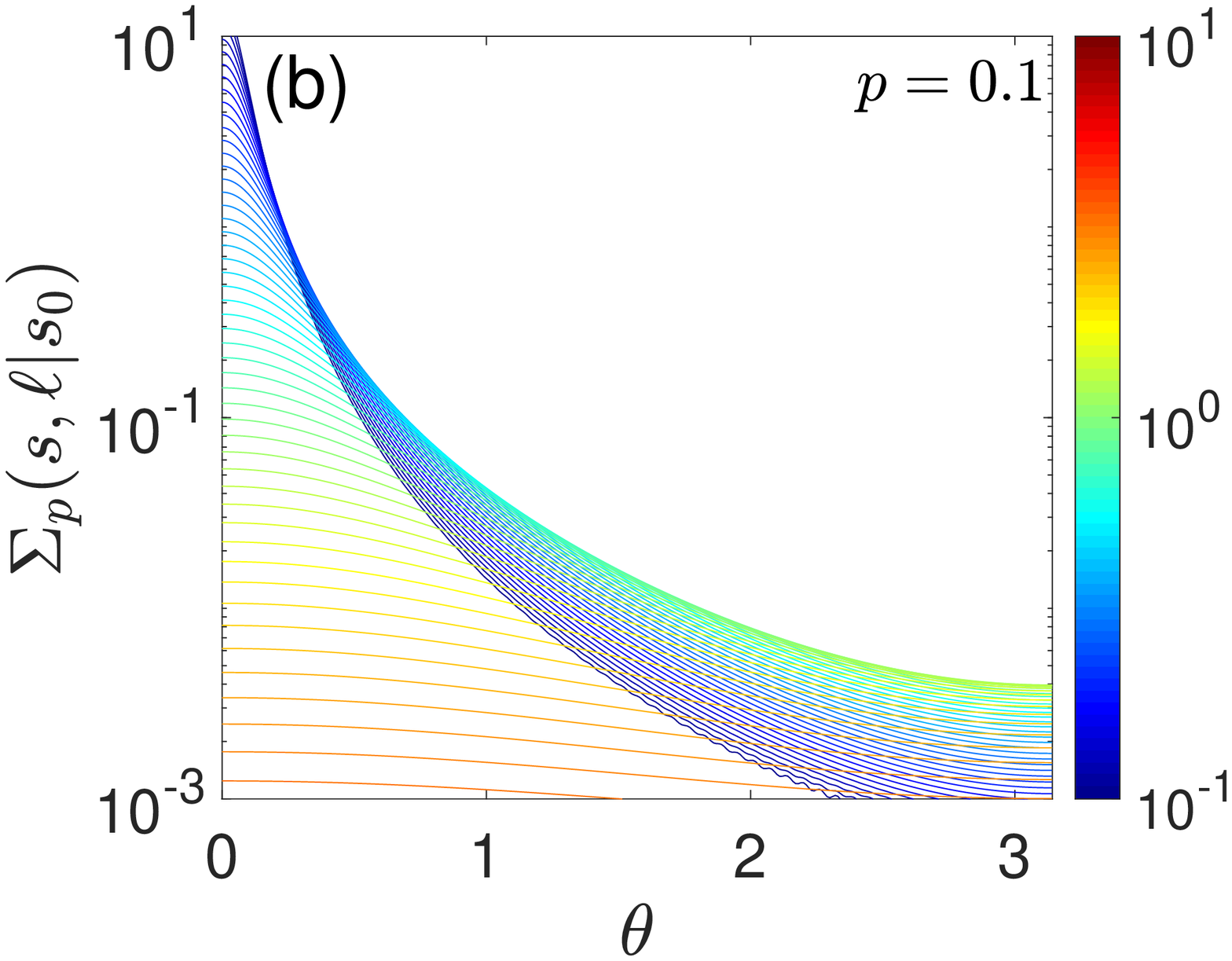} % Sigma_shell_p01.eps}
\end{center}
\caption{
The surface hopping propagator $\Sigma_p(\s,\ell|\s_0)$, given by
Eq. (\ref{eq:Sigma_shell}), on a spherical target of radius $R$,
surrounded by an outer reflecting sphere of radius $L$, for $p = 0$
{\bf (a)} and $p = 0.1$ {\bf (b)}, and 64 values of $\ell$,
logarithmically spaced in the range from $10^{-1}$ (blue curves) to
$10^1$ (red curves), with $L = 10R$, $D = 1$, $\s_0 = (0,0,R)$ (the
North pole), and $\s = (R\sin \theta,0,R\cos\theta)$ (i.e., $\theta$
is the angle between $\s_0$ and $\s$).  The series in
Eq. (\ref{eq:Sigma_shell}) is truncated above $n = 100$. }
\label{fig:Sigma_shell}
% [Sigma, ell,s] = A_localtime_Sigma_fig;
% Exported as 20 x 16
\end{figure}

Finally, Fig. \ref{fig:Sigma_shell} illustrates the surface hopping
propagator on a spherical target of radius $R$, surrounded by an outer
reflecting sphere of radius $L$.  This setting is qualitatively in
between the interior and the exterior of a ball.  On one hand, as this
domain is bounded and diffusion is recurrent, the surface hopping
propagator evolves towards the uniform density $1/(4\pi R^2)$, as for
the interior case.  On the other hand, the outer reflecting sphere is
located relatively far from the target and thus allows for long
trajectories, as for the exterior case.  For $p = 0$, the panel (a) of
Fig. \ref{fig:Sigma_shell} resembles the panel (a) of
Fig. \ref{fig:Sigma_ball}, even so diffusion occurs in different
regions in these two settings.  In contrast, when $ p = 0.1$, the
panel (b) of Fig. \ref{fig:Sigma_shell} is much closer to the panel
(d) of Fig. \ref{fig:Sigma_ball}.  In fact, in both cases, the
particle is allowed to undertake long trajectories between successive
encounters with the target.

\section{Discussion and Conclusion}
\label{sec:discussion}

In this paper, we investigated the properties of the surface hopping
propagator $\Sigma_p(\s,\ell|\s_0)$ recently introduced in
\cite{Grebenkov20}.  This is a conceptually new quantity that
describes bulk-diffusion-mediated exploration of a surface.  In
contrast to former works
\cite{Chechkin09,Chechkin11,Chechkin12,Berezhkovskii15,Berezhkovskii17},
which relied on coupled bulk-surface diffusion equations and aimed to
characterize the position of the particle on a surface after some
physical time $t$, here we operate with the boundary local time
$\ell$, which is a proxy of the number of encounters with that
surface.  The surface hopping propagator turns out to be {\it dual} to
the conventional propagator $G_q(\x,t|\x_0)$.  In fact, as
$G_q(\x,t|\x_0)$ characterizes displacements between bulk points
$\x_0$ and $\x$ in physical time $t$ (i.e., after a number of bulk
jumps), $\Sigma_p(\s,\ell|\s_0)$ characterizes effective displacements
between boundary points $\s_0$ and $\s$ in boundary local time $\ell$
(i.e., after a number of reflections on the boundary).  While
$G_q(\x,t|\x_0)$ is the semi-group of the Laplace operator $-\Delta$
(acting in the bulk), $\Sigma_p(\s,\ell|\s_0)$ is the semi-group of
the Dirichlet-to-Neumann operator $\M_p$ (acting on the boundary).  In
this light, the spectral expansion (\ref{eq:Sigma_M}) is dual to the
spectral expansion for the conventional propagator:
\begin{equation}  \label{eq:Gq_eigen}
G_q(\x,t|\x_0) = \sum\limits_n [u_n^{(q)}(\x_0)]^* \, u_n^{(q)}(\x) \, e^{-\lambda_n^{(q)} t} ,
\end{equation}
where $\lambda_n^{(q)}$ and $u_n^{(q)}(\x)$ are the eigenvalues and
$L_2(\Omega)$-normalized eigenfunctions of the diffusion operator
$-D\Delta$:
\begin{subequations}
\begin{align}
-D \Delta u_n^{(q)} & = \lambda_n^{(q)} \, u_n^{(q)}  \qquad (\x \in \Omega), \\  \label{eq:un_Robin}
\partial_{\n} u_n^{(q)} + q \, u_n^{(q)} & = 0 \qquad (\x \in \pa) ,
\end{align}
\end{subequations}
where we highlighted the dependence on the reactivity parameter $q$
through the Robin boundary condition (\ref{eq:un_Robin}).  Similarity
and duality of Eqs. (\ref{eq:Sigma_M}, \ref{eq:Gq_eigen}) are
remarkable.  We recall that the spectral expansion (\ref{eq:Gq_eigen})
is valid for a bounded domain $\Omega$, whereas the spectral expansion
(\ref{eq:Sigma_M}) is valid for a bounded boundary $\pa$.  As a
consequence, Eq. (\ref{eq:Sigma_M}) seems to be more general, as it is
also valid when $\Omega$ is the exterior of a bounded domain, for
which Eq. (\ref{eq:Gq_eigen}) is not applicable anymore.  In spite of
this considerable advantage, the Dirichlet-to-Neumann operator and its
eigenbasis were not earlier employed to describe diffusion-influenced
reactions and other diffusion-mediated surface phenomena.  The present
paper, along with Refs. \cite{Grebenkov20,Grebenkov19}, aim to shift
the theoretical description of these phenomena towards a new
fundamental ground.  Using the formulas derived in this paper, one can
access directly not only the surface hopping propagator, but all the
related quantities, including the full propagator, first-passage time
distribution, the survival probability and the reaction rate.

\begin{acknowledgments}
The author acknowledges a partial financial support from the Alexander von
Humboldt Foundation through a Bessel Research Award. 
\end{acknowledgments}

\appendix
\section{Circular annulus}
\label{sec:annulus}

Here, we consider a circular annulus between two concentric circles of
radii $R < L$: $\Omega = \{\x\in\R^2 ~:~ R < |\x| < L\}$.  There are
four possible combinations of surface reactivity: (i) both circles are
reactive, (ii) the inner circle is reflecting while the outer circle
is reactive, (iii) the inner circle is reactive while the outer circle
is reflecting, and (iv) both circles are reflecting.  As surface
reaction is not possible in the last case, it is excluded.  The first
case corresponds directly to our general setting when the whole
boundary is reactive.  However, as this case involves two disjoint
parts of the boundary (the inner and the outer circles), its analysis
is the most complicated.  For this reason, we start with the case
(ii), then briefly discuss the case (iii), and finally give the
solution for the case (i).

\subsection{Reactive outer circle}

In order to determine the spectrum of the Dirichlet-to-Neumann
operator associated with the outer circle $\Gamma = \{\x\in \R^2 ~:~
|\x| = L\}$, one needs to solve the mixed boundary value problem
(\ref{eq:u_def2}).  Its general solution can be searched in polar
coordinates $(r,\phi)$ as:
\begin{equation}  \label{eq:w_annulus}
w(r,\phi) = \sum\limits_{n=-\infty}^\infty c_n \, e^{in\phi} \, g_n(r),
\end{equation}
where $c_n$ are unknown coefficients (to be fixed by the boundary
condition), 
\begin{equation}  \label{eq:gn_ND}
g_n(r) = \frac{K'_n(\alpha R) I_n(\alpha r) - I'_n(\alpha R) K_n(\alpha r)}
{K'_n(\alpha R) I_n(\alpha L) - I'_n(\alpha R) K_n(\alpha L)} \,.
\end{equation}
are radial functions with $\alpha = \sqrt{p/D}$, prime denotes the
derivative with respect to the argument, and $I_n(z)$ is the modified
Bessel functions of the first kind.  One can easily check that
$g'_n(R) = 0$ by construction.  For convenience, we have chosen a
particular normalization $g_n(L) = 1$.  As the normal derivative is
equal to the radial derivative, the action of the Dirichlet-to-Neumann
operator $\M_p$ does not affect the angular part.  In other words, the
rotational symmetry of this domain implies that Fourier harmonics are
the eigenfunctions of $\M_p$, defined on the outer circle $\Gamma$:
\begin{equation}  \label{eq:vn_annulus}
v_n(s) = \frac{e^{in s/L}}{\sqrt{2\pi L}}  \qquad (n \in \Z), 
\end{equation}
where the curvilinear coordinate $s$ is related to the polar angle
$\phi$ as $s = \phi L$.  The associated eigenvalues are
\begin{equation}  \label{eq:mu_annulus}
\mu_n^{(p)} = g'_n(L)  \qquad (n \in \Z).
\end{equation}
The eigenfunctions do not depend on $p$, whereas the eigenvalues are
twice degenerate, except for $n = 0$.  Here, the index $n$ runs over
all integer numbers.  In the limit $p\to 0$, one gets
\begin{equation}  \label{eq:mu_p0_annulus}
\mu_n^{(0)} = \frac{|n|}{L} \, \frac{1 - (R/L)^{2|n|}}{1 + (R/L)^{2|n|}} \,.
\end{equation}
The spectral decomposition (\ref{eq:Sigma_M}) of the surface hopping
propagator reads
\begin{equation}    \label{eq:Sigmap_annulus}
\Sigma_p(s,\ell|s_0) = \sum\limits_{n=-\infty}^\infty \frac{e^{in(s_0-s)/L}}{2\pi L} 
\exp\bigl( - \mu_n^{(p)} \ell \bigr).
\end{equation}

To access the full propagator, one also needs to compute
$V_n^{(p)}(\x_0)$ from Eq. (\ref{eq:Vnp}).  Using the summation
formulas from \cite{Grebenkov19g}, the Laplace-transformed propagator
$\tilde{G}_\infty(\x,p|\x_0)$ and thus $\tilde{j}_\infty(\s,p|\x_0)$
for a circular annulus with Dirichlet boundary condition on the outer
circle and Neumann boundary condition on the inner circle read
\begin{align}   \label{eq:Ginf_annulus}
\tilde{G}_\infty(\x,p|\x_0) &= \frac{1}{2\pi D} \sum\limits_{n=-\infty}^\infty e^{in(\phi-\phi_0)} \, g_n(r_0) \\   \nonumber
& \times \bigr[K_n(\alpha L) I_n(\alpha r) - I_n(\alpha L) K_n(\alpha r)\bigr] \,, \\  \label{eq:jinf_annulus}
\tilde{j}_\infty(s,p|\x_0) &= \frac{1}{2\pi L} \sum\limits_{n=-\infty}^\infty e^{in(\phi-\phi_0)} \, g_n(r_0) ,
\end{align}
where $\x = (r,\phi)$ and $\x_0 = (r_0,\phi_0)$ in polar coordinates,
$R \leq r_0 \leq r \leq L$, $s = \phi L$, and we used the Wronskian
\begin{equation} \label{eq:Wronskian_2d}
I'_n(z) K_n(z) - K'_n(z) I_n(z) = \frac{1}{z} \,.  
\end{equation}
The projection of $\tilde{j}_\infty(\s,p|\x_0)$ onto an eigenfunction
$v_{n}(s)$ from Eq. (\ref{eq:vn_annulus}) reads then
\begin{equation}  \label{eq:Vnp_annulus}
V_{n}^{(p)}(\x_0) = v_{n}(\phi_0) \, g_n(r_0) \,.
\end{equation}
The orthogonality of Fourier harmonics to a constant function reduces
Eq. (\ref{eq:U2p}) to
\begin{equation}  \label{eq:Up_annulus}
\tilde{U}(\ell,p|\x_0) =  g_0(r_0)\, \exp(-\mu_0^{(p)}\ell)  ,
\end{equation}
from which Eq. (\ref{eq:Hpsi}) gives the Laplace-transformed
probability density of the reaction time as
\begin{equation}  \label{eq:Hp_annulus}
\tilde{H}_q(p|\x_0) = g_0(r_0) \, \frac{1}{1 + \mu_0^{(p)}/q} \,.
\end{equation}

\subsection{Interior of a disk}

In the limit $R\to 0$, the inner boundary shrinks to a point, and one
gets the solution for the interior of a disk of radius $L$: $\Omega =
\{ \x\in\R^2~: ~|\x| < L\}$ (see also \cite{Grebenkov19}).  The radial
functions become
\begin{equation}  \label{eq:gn_diskI}
g_n(r) = \frac{I_n(r\sqrt{p/D})}{I_n(L\sqrt{p/D})} \,,
\end{equation}
while the eigenvalues and eigenfunctions of the Dirichlet-to-Neumann
operator are still given by Eqs. (\ref{eq:vn_annulus},
\ref{eq:mu_annulus}).  Other expressions are also valid; in particular, 
Eqs. (\ref{eq:Ginf_annulus}, \ref{eq:jinf_annulus}) are applicable.
  
At $p = 0$, Eq. (\ref{eq:mu_p0_annulus}) yields $\mu_n^{(0)} = |n|/L$
and thus
\begin{equation}  \label{eq:Sigma_disk}
\Sigma_0(s,\ell|s_0) = \frac{1 - e^{-2\ell/L}}{2\pi L\bigl(1 - 2\cos(\frac{s-s_0}{L}) e^{-\ell/L} + e^{-2\ell/L}\bigr)} \,.
\end{equation}
As expected, this propagator evolves from the Dirac distribution
$\delta(s-s_0)$ at $\ell = 0$ to the uniform distribution $1/(2\pi L)$
as $\ell\to\infty$.  Setting $\rho = L e^{-\ell/L}$, one can recognize
in this form the Poisson kernel in the disk of radius $L$.  The
Poisson kernel describes the harmonic measure density on the disk,
i.e., the probability density of the first arrival onto the circle of
radius $R$ at point $(L,s/L)$ for Brownian motion started from a point
$(\rho,s_0/L)$ (written in polar coordinates).  As in the planar case
discussed in Sec. \ref{sec:half}, the distribution of the position of
the diffusing particle at the boundary local time $\ell$ (i.e., after
a prescribed number of encounters with the reflecting circle) is
identical to the distribution of the first arrival point on the fully
absorbing circle, where the boundary local time $\ell$ determines the
starting point in the latter setting.  Figure
\ref{fig:Sigma_disk}(a,b) illustrates the behavior of the surface
hopping propagator.

\subsection{Reactive inner circle}

We briefly discuss the case (iii) when the inner circle is reactive
and surrounded by a reflecting outer circle.  This is a typical
setting of a small reactive target confined in a domain surrounded by
an outer reflecting boundary \cite{Grebenkov18}.  Here, we search for
the spectrum of the Dirichlet-to-Neumann operator associated with the
inner circle: $\Gamma = \{ \x\in\R^2 ~:~ |\x| = R\}$.  Repeating the
above construction step by step, one realizes that the eigenfunctions
are the Fourier harmonics on the inner circle
\begin{equation}  \label{eq:vn_annulus2}
v_n(s) = \frac{e^{in s/R}}{\sqrt{2\pi R}}  \qquad (n \in \Z), 
\end{equation}
where the curvilinear coordinate $s$ is related to the polar angle
$\phi$ as $s = \phi R$.  The associated eigenvalues are
\begin{equation}  \label{eq:mu_annulus2}
\mu_n^{(p)} = - g'_n(R)  \qquad (n \in \Z),
\end{equation}
where sign minus appears due to the direction of the normal
derivative, $\partial_{\n} = -\partial_r$, and
\begin{equation}  \label{eq:gn_DN}
g_n(r) = \frac{K'_n(\alpha L) I_n(\alpha r) - I'_n(\alpha L) K_n(\alpha r)}
{K'_n(\alpha L) I_n(\alpha R) - I'_n(\alpha L) K_n(\alpha R)} 
\end{equation}
are the radial functions satisfying $g_n(R) = 1$ and $g'_n(L) = 0$.
In the limit $p\to 0$, one gets
\begin{equation}
\mu_n^{(0)} = \frac{|n|}{R} \, \frac{1 - (R/L)^{2|n|}}{1 + (R/L)^{2|n|}} \,.
\end{equation}

The spectral decomposition (\ref{eq:Sigma_M}) of the surface hopping
propagator reads
\begin{equation}    \label{eq:Sigmap_annulus2}
\Sigma_p(s,\ell|s_0) = \sum\limits_{n=-\infty}^\infty \frac{e^{in(s_0-s)/R}}{2\pi R} 
\exp\bigl( - \mu_n^{(p)} \ell \bigr).
\end{equation}

The Laplace-transformed propagator $\tilde{G}_\infty(\x,p|\x_0)$ and
thus $\tilde{j}_\infty(\s,p|\x_0)$ for a circular annulus with
Dirichlet boundary condition on the inner circle and Neumann boundary
condition on the outer circle read
\begin{align}   \label{eq:Ginf_annulus2}
\tilde{G}_\infty(\x,p|\x_0) &= \frac{1}{2\pi D} \sum\limits_{n=-\infty}^\infty e^{in(\phi-\phi_0)} \, g_n(r_0) \\   \nonumber
& \times \bigr[K_n(\alpha R) I_n(\alpha r) - I_n(\alpha R) K_n(\alpha r)\bigr] \,, \\  \label{eq:jinf_annulus2}
\tilde{j}_\infty(s,p|\x_0) &= \frac{1}{2\pi R} \sum\limits_{n=-\infty}^\infty e^{in(\phi-\phi_0)} \, g_n(r_0) ,
\end{align}
where $\x = (r,\phi)$, $\x_0 = (r_0,\phi_0)$, $R \leq r \leq r_0 \leq
L$, $s = \phi R$.  In turn, Eqs. (\ref{eq:Vnp_annulus},
\ref{eq:Up_annulus}, \ref{eq:Hp_annulus}) remain unchanged.  Some
first-passage properties in this setting were studied in
\cite{Grebenkov20}.

\subsection{Exterior of a disk}
\label{sec:diskE}

In the limit $L\to\infty$, the outer boundary is pushed away to
infinity, and one deals with diffusion in the exterior of a disk of
radius $R$: $\Omega = \{\x\in\R^2 ~:~ |\x| > R\}$.  In this limit, the
radial functions from Eq. (\ref{eq:gn_DN}) are reduced to
\begin{equation}  \label{eq:gn_diskE}
g_n(r) = \frac{K_n(r\sqrt{p/D})}{K_n(R\sqrt{p/D})} \,.
\end{equation}
The eigenvalues and eigenfunctions of the Dirichlet-to-Neumann
operator are still given by Eqs. (\ref{eq:mu_annulus2},
\ref{eq:vn_annulus2}), with $g_n(r)$ from Eq. (\ref{eq:gn_diskE}).
Other earlier expressions are as well applicable; in particular,
Eqs. (\ref{eq:Ginf_annulus2}, \ref{eq:jinf_annulus2}) are valid.
Figure \ref{fig:Sigma_disk}(c,d) illustrates the behavior of the
surface hopping propagator.

The exterior of a disk presents a convenient example to illustrate
subtle points of recurrent diffusion outside a planar bounded domain.
As the radial functions $g_n(r)$ in Eq. (\ref{eq:gn_diskE}) for $p >
0$ vanish exponentially fast in the limit $r\to \infty$, a general
solution $w(r,\phi)$ of the modified Helmholtz equation
(\ref{eq:u_def2}) also vanishes, in agreement with the regularity
condition.  In the case $p = 0$, the radial functions for $n \ne 0$
become $g_n(r) = (R/r)^{|n|}$ and vanish again.  However, the limit of
$g_0(r)$ as $p\to 0$ is equal to $1$ that does not vanish at infinity,
thus violating the regularity condition.  This is a consequence of the
simple fact that the rotationally invariant Laplace equation in the
plane, $w'' + \frac{1}{r} w' = 0$, has a general solution $c_1 +
c_2\ln r$ that does not vanish as $r\to\infty$, except for the trivial
choice $c_1 = c_2 = 0$.  This is a well-known problem for such planar
domains, for which, in particular, there is no steady-state reaction
rate \cite{Torney83,Grebenkov19j}.  In the remaining part of the
paper, we do not discuss this subtle case.

\subsection{Both reactive circles}

When both circles are reactive, one needs to consider the
Dirichlet-to-Neumann operator on the whole boundary composed of two
disjoint circles: $\Gamma_1 = \{ \x\in\R^2 ~: ~|\x| = R\}$ and
$\Gamma_2 = \{ \x\in\R^2 ~: ~|\x| = L\}$.  A general solution of
Eq. (\ref{eq:u_def}) can be searched in the form
\begin{equation}  \label{eq:w_DDcase}
w(r,\phi) = \sum\limits_{n=-\infty}^\infty \bigl(c_{n,1}\, g_{n,1}(r) + c_{n,2}\, g_{n,2}(r)\bigr) e^{in\phi},
\end{equation}
where the unknown coefficients $c_{n,1}$ and $c_{n,2}$ are set by
boundary conditions, and the radial functions
\begin{align*}
g_{n,1}(r) & = \frac{K_n(\alpha L) I_n(\alpha r) - I_n(\alpha L) K_n(\alpha r)}
{K_n(\alpha L) I_n(\alpha R) - I_n(\alpha L) K_n(\alpha R)} \,, \\
g_{n,2}(r) & = \frac{K_n(\alpha R) I_n(\alpha r) - I_n(\alpha R) K_n(\alpha r)}
{K_n(\alpha R) I_n(\alpha L)  - I_n(\alpha R) K_n(\alpha L)} 
\end{align*}
satisfy $g_{n,1}(L) = 0$, $g_{n,1}(R) = 1$, and $g_{n,2}(R) = 0$,
$g_{n,2}(L) = 1$ for convenience (other linear combinations of
$I_n(\alpha r)$ and $K_n(\alpha r)$ could also be used).  Note that
$g_{n,1}(r)$ monotonously decreases, whereas $g_{n,2}(r)$
monotonously increases on the interval $(R,L)$.

As the inner and outer circles are concentric, one can expect that an
eigenfunction of the operator $\M_p$ can be written as 
\begin{equation}
v_n^{(p)}(s) = \left\{ \begin{array}{l l} a_n^{(p)} \, e^{in \phi} & \quad s \in \Gamma_1 ~~ (\phi = s/R), \\  
b_n^{(p)} \, e^{in \phi}  & \quad s \in \Gamma_2 ~~ (\phi = s/L), \\  \end{array} \right. 
\end{equation}
where $a_n^{(p)}$ and $b_n^{(p)}$ are some coefficients.  In fact, as
the space of functions on the whole boundary $\pa = \Gamma_1 \cup
\Gamma_2$ is the direct product of spaces of functions on the inner
($\Gamma_1$) and outer ($\Gamma_2$) circles, the eigenfunction
$v_n^{(p)}(s)$ can be thought of being composed of two components.
Here, these two components are proportional to $e^{in\phi}$ due to the
rotational symmetry (note that this claim can be shown rigorously by
representing each component as a Fourier series and then using the
orthogonality of Fourier harmonics).  Substituting such $v_n^{(p)}(s)$
as the Dirichlet condition to Eq. (\ref{eq:u_def}), one gets its
solution as
\begin{equation}  \label{eq:w_DDcase1}
w(r,\phi) = \bigl(a_n^{(p)}\,  g_{n,1}(r) + b_n^{(p)} \, g_{n,2}(r)\bigr) e^{in\phi},
\end{equation}
while its normal derivative on $\pa$ reads
\begin{align}
\M_p v_n & = (\partial_n w)_{|\pa} \\  \nonumber
& = \left\{ \begin{array}{l l} - (a_n^{(p)}\, g'_{n,1}(R) + b_n^{(p)}\, g'_{n,2}(R)) e^{in \phi} & \quad s \in \Gamma_1, \\  
(a_n^{(p)} \, g'_{n,1}(L) + b_n^{(p)} \, g'_{n,2}(L)) e^{in \phi}  & \quad s \in \Gamma_2. \\  \end{array} \right. 
\end{align}
If the right-hand side is proportional to $v_n^{(p)}$, then
$v_n^{(p)}$ is indeed an eigenfunction of $\M_p$.  In other words, we
get two equations
\begin{equation}  
\left\{ \begin{array}{l} - \bigl(a_n^{(p)}\, g'_{n,1}(R) + b_n^{(p)}\, g'_{n,2}(R)\bigr) = \mu_n^{(p)}\, a_n^{(p)}  , \\
\bigl(a_n^{(p)} \, g'_{n,1}(L) + b_n^{(p)} \, g'_{n,2}(L)\bigr) = \mu_n^{(p)}\, b_n^{(p)} , \\  \end{array} \right. 
\end{equation}
where the proportionality coefficient $\mu_n^{(p)}$ is the associated
eigenvalue, and we used the particular form of functions $g_{n,1}(r)$
and $g_{n,2}(r)$.  These equations can be written in a matrix form as
\begin{equation}  \label{eq:eigen_2x2}
\left( \begin{array}{ll} - g'_{n,1}(R) & - g'_{n,2}(R) \\  g'_{n,1}(L) & g'_{n,2}(L) \\  \end{array} \right)
\left( \begin{array}{ll} a_n^{(p)} \\  b_n^{(p)} \\  \end{array} \right) = \mu_n^{(p)} 
\left( \begin{array}{ll} a_n^{(p)} \\  b_n^{(p)} \\  \end{array} \right) .
\end{equation}
Solving this eigenvalue problem for the $2\times 2$ matrix on the
left-hand side, one determines the eigenvalue $\mu_n^{(p)}$, as well
as one of the coefficients (e.g., $a_n^{(p)}$).  The other coefficient
(e.g., $b_n^{(p)}$) is fixed by imposing the $L_2(\pa)$-normalization
of the eigenfunction $v_n^{(p)}$.

The Wronskian (\ref{eq:Wronskian_2d}) yields
\begin{equation}  \label{eq:ggA}
g'_{n,1}(L) = \frac{A}{L} \,, \qquad g'_{n,2}(R) = - \frac{A}{R} \, ,
\end{equation}
where
\begin{equation}
A = \frac{1}{K_n(\alpha L) I_n(\alpha R) - I_n(\alpha L) K_n(\alpha R)} \,.
\end{equation}
The eigenvalue $\mu_n^{(p)}$ is then obtained as a solution of the
quadratic equation 
\begin{equation}
\mu^2 + \mu \bigl( g'_{n,1}(R) - g'_{n,2}(L) \bigr) + B = 0 ,
\end{equation}
where 
\begin{align}  \nonumber
B & = - g'_{n,1}(R) g'_{n,2}(L) + g'_{n,1}(L) g'_{n,2}(R) \\
& = - \alpha^2 \frac{K'_n(\alpha L) I'_n(\alpha R) - I'_n(\alpha L) K'_n(\alpha R)}{K_n(\alpha L) I_n(\alpha R) - I_n(\alpha L) K_n(\alpha R)} \,.
\end{align}
With the help of Eq. (\ref{eq:ggA}), it is easy to check that the
determinant of this equation is positive, so that there are two real
roots.  Moreover, as $B > 0$, both roots are positive:
\begin{align}  \label{eq:mu_shell_DD}
\mu_{n,\pm}^{(p)} & = \frac12 \biggl( g'_{n,2}(L) - g'_{n,1}(R)  \\  \nonumber
& \pm \sqrt{(g'_{n,1}(R) + g'_{n,2}(L))^2 - 4 g'_{n,1}(L) g'_{n,2}(R)} \biggr) .
\end{align}
As a consequence, for each index $n$, there are two distinct
eigenmodes.  For each of them, the coefficients are determined by the
corresponding eigenvalue.  To avoid round-off errors in practical
implementation, it is convenient to use slightly different (but
formally equivalent) representation for $+$ and $-$ modes.  In fact,
we set
\begin{equation}  \label{eq:ab_shell_DD}
a_{n,+}^{(p)} = - c_{n,+}^{(p)} (g'_{n,2}(L) - \mu_{n,+}^{(p)}), \qquad b_{n,+}^{(p)} = c_{n,+}^{(p)} \, g'_{n,1}(L) \,,
\end{equation}
where $c_{n,+}^{(p)}$ is fixed by the normalization of the
eigenfunction:
\begin{align*}  
1 & = \int\limits_{\pa} ds \, |v_{n,+}^{(p)}(s)|^2 \\
& = [c_{n,+}^{(p)}]^2 \left(2\pi R [g'_{n,2}(L) - \mu_{n,+}^{(p)}]^2 + 2\pi L [g'_{n,1}(L)]^2 \right).
\end{align*}
In turn, for the mode with $\mu_{n,-}^{(p)}$, one can use
\begin{equation}  \label{eq:ab_shell_DD-}
a_{n,-}^{(p)} = c_{n,-}^{(p)} \, g'_{n,2}(R) , \qquad b_{n,-}^{(p)} = - c_{n,-}^{(p)} (g'_{n,1}(R) + \mu_{n,-}^{(p)}) \,,
\end{equation}
with
\begin{align*}  
1 = [c_{n,-}^{(p)}]^2 \left( 2\pi R \, [g'_{n,2}(R)]^2 + 2\pi L [g'_{n,1}(R) + \mu_{n,-}^{(p)}]^2\right).
\end{align*}

In the limit $p\to 0$, one has
\begin{align*}
g'_{n,1}(R) & \to - \frac{|n|}{R} \, \frac{1+(R/L)^{2|n|}}{1-(R/L)^{2|n|}} \,, \\
g'_{n,1}(L) & \to - \frac{|n|}{L} \, \frac{2(R/L)^{|n|}}{1-(R/L)^{2|n|}} \,, \\
g'_{n,2}(R) & \to \frac{|n|}{R} \, \frac{2(R/L)^{|n|}}{1-(R/L)^{2|n|}} \,, \\
g'_{n,2}(L) & \to \frac{|n|}{L} \, \frac{1+(R/L)^{2|n|}}{1-(R/L)^{2|n|}} \,, 
\end{align*}
so that
\begin{align}
\mu_{n,\pm}^{(0)} & = \frac{|n|}{2} \, \frac{1+\gamma_n}{1-\gamma_n} \left(\frac{1}{L} + \frac{1}{R}  \right. \\  \nonumber
& \left. \pm \sqrt{\biggl(\frac{1}{L} + \frac{1}{R}\biggr)^2 - \frac{4(1-\gamma_n)^2}{LR(1+\gamma_n)^2}} \right)  ,
\end{align}
where $\gamma_n = (R/L)^{2|n|}$.  In the case $n = 0$, one can take
the limit $n\to 0$ to get
\begin{equation}
\mu_{0,-}^{(0)} = 0,  \qquad \mu_{0,+}^{(0)} = \frac{1/L + 1/R}{\ln(L/R)} \,.
\end{equation}
We also get
\begin{align*}
a_{0,-}^{(0)} & = b_{0,-}^{(0)} = \frac{1}{\sqrt{2\pi(R+L)}} \,, \\
a_{0,+}^{(0)} & = \frac{1}{\sqrt{2\pi R(1+R/L)}} , \qquad b_{0,+}^{(0)} = \frac{-1}{\sqrt{2\pi L(1 + L/R)}}  \,.
\end{align*}

Using the asymptotic behavior of the modified Bessel functions, one
can check that 
\begin{align*}
\lim\limits_{R\to 0} \mu_{n,-}^{(p)} & = \sqrt{p/D} \,\frac{I'_n(L\sqrt{p/D})}{I_n(L\sqrt{p/D})} \,, \\
\lim\limits_{R\to 0} \mu_{n,+}^{(p)} & = +\infty \,, \\
\lim\limits_{L\to \infty} \mu_{n,-}^{(p)} & = - \sqrt{p/D} \,\frac{K'_n(R\sqrt{p/D})}{K_n(R\sqrt{p/D})} \,,  \\
\lim\limits_{L\to \infty} \mu_{n,+}^{(p)} & = \sqrt{p/D} \,,  \\
\end{align*}
where we used that $K'_n(z)/K_n(z) \leq - 1$.  In the limit $R\to 0$,
$\mu_{n,-}^{(p)}$ approach the eigenvalues of $\M_p$ for the interior
of a disk of radius $L$, whereas $\mu_{n,+}^{(p)}$ diverge and thus do
not contribute.  The opposite limit $L\to \infty$ is more subtle:
$\mu_{n,+}^{(p)}$ approach the eigenvalues of $\M_p$ for the exterior
of a disk of radius $R$; however, $\mu_{n,-}^{(p)}$ accumulate near
$\sqrt{p/D}$.

Finally, the Dirichlet propagator in the Laplace domain is
\begin{subequations}
\begin{align}   \label{eq:Ginf_annulus3}
& \tilde{G}_\infty(\x,p|\x_0) = \frac{1}{2\pi D} \sum\limits_{n=-\infty}^\infty e^{in(\phi-\phi_0)} \, g_{n,2}(r_0) \\   \nonumber
& \times \bigr[K_n(\alpha L) I_n(\alpha r) - I_n(\alpha L) K_n(\alpha r)\bigr] \quad  (r_0 \leq r),  \\ 
& \hskip 19mm = \frac{1}{2\pi D} \sum\limits_{n=-\infty}^\infty e^{in(\phi-\phi_0)} \, g_{n,1}(r_0) \\   \nonumber
& \times \bigr[K_n(\alpha R) I_n(\alpha r) - I_n(\alpha R) K_n(\alpha r)\bigr] \quad  (r \leq r_0),
\end{align}
\end{subequations}
where $\x = (r,\phi)$ and $\x_0 = (r_0,\phi_0)$.  The probability flux
density reads then
\begin{subequations}
\begin{align}
\tilde{j}_\infty(s,p|\x_0) &= \frac{1}{2\pi R} \sum\limits_{n=-\infty}^\infty e^{in(\phi-\phi_0)} \, g_{n,1}(r_0)  \quad (s\in\Gamma_1) , \\
&= \frac{1}{2\pi L} \sum\limits_{n=-\infty}^\infty e^{in(\phi-\phi_0)} \, g_{n,2}(r_0)  \quad (s\in\Gamma_2) .
\end{align}
\end{subequations}
As a consequence, one gets
\begin{equation}
V_n^{(p)}(\x_0) = \bigl(a_n^{(p)} \, g_{n,1}(r_0) + b_n^{(p)} \, g_{n,2}(r_0) \bigr) e^{in\phi_0} 
\end{equation}
that gives access to the full propagator $P(\x,\ell,t|\x_0)$.

\subsection{Cylindrical domains}

The above analysis can also be extended to cylindrical domains.  Let
us first consider an infinite cylinder, $\Omega = \Omega_0 \times \R$,
where $\Omega_0$ is a disk of radius $L$.  As the boundary $\pa$ is
unbounded, the spectrum of the Dirichlet-to-Neumann operator $\M_p$ is
not discrete anymore.  Nevertheless, the symmetries of this domain
admit the separation of variables and allow for getting
``eigenfunctions'' and ``eigenvalues'' in cylindrical coordinates as
\begin{subequations}
\begin{align}
v_{nk}(\phi,z) & = \frac{e^{in\phi + ikz}}{2\pi \, \sqrt{L}}   \qquad (n\in \Z,~ k\in \R) , \\
\mu_{nk}^{(p)} & = \alpha \, \frac{I'_n(\alpha L)}{I_n(\alpha L)} \,, \quad \alpha = \sqrt{p/D + k^2} .
\end{align}
\end{subequations}
The surface hopping propagator reads then
\begin{equation}
\Sigma_p(\s,\ell|\s_0) = \sum\limits_{n=-\infty}^{\infty} \frac{e^{in(\phi_0-\phi)}}{2\pi L} 
 \int\limits_\R \frac{dk}{2\pi} e^{ik(z_0-z)} \, e^{-\mu_{nk}^{(p)}\ell} ,
\end{equation}
where $\s = (L,\phi,z)$ and $\s_0 = (L,\phi_0,z_0)$ in cylindrical
coordinates.  If $\Omega_0$ is a circular annulus, one has to use the
appropriate radial function $g_n(r)$, with $\alpha = \sqrt{p/D +
k^2}$.  Other related quantities can also be obtained.

When the cylinder is finite, $\Omega = \Omega_0 \times (0,b)$, the
spectrum of $\M_p$ is discrete again, but the analysis is more subtle.
In fact, as in the case of a circular annulus, different combinations
of reactivity patterns are possible: all the boundary is reactive;
only the lateral boundary is reactive but the top and bottom disks are
reflecting; only the top disk is reactive but the remaining boundary
is reflecting; etc.  When only one part of the boundary is reactive,
the analysis is rather simple.  For instance, if only the lateral
boundary is reactive, the eigenfunctions and eigenvalues are 
\begin{subequations}
\begin{align}
v_{nk}(\phi,z) & = \frac{e^{in\phi}}{\sqrt{2\pi L}} \, \frac{\sqrt{2-\delta_{n,0}}}{\sqrt{b}} \cos(\pi k z/b) , \\
\mu_{nk}^{(p)} & = \alpha \, \frac{I'_n(\alpha L)}{I_n(\alpha L)} \,, \quad \alpha = \sqrt{p/D + (\pi k/b)^2} ,
\end{align}
\end{subequations}
with $n\in\Z$ and $k=0,1,2,\ldots$.  However, the analysis is more
involved when the whole boundary is reactive.

\section{Spherical shell}
\label{sec:shell}

In three dimensions, one can consider a spherical shell between two
concentric spheres of radii $R < L$: $\Omega = \{\x\in\R^3 ~:~ R <
|\x| < L\}$, with three combinations of boundary conditions.  As the
analysis is rather similar to the two-dimensional setting, the results
are presented in a concise form.

\subsection{Reactive outer sphere}

As previously, we start with the case of the Dirichlet-to-Neumann
operator $\M_p$ associated with the reactive outer sphere $\Gamma = \{
\x\in\R^3 ~:~ |\x| = L\}$.  The rotational invariance implies that the
eigenfunctions of $\M_p$ are the (normalized) spherical harmonics,
\begin{equation}  \label{eq:vn_shell}
v_{nm}(\s) = \frac{1}{L} \, Y_{mn}(\theta,\phi)  \quad (n=0,1,2,\ldots, ~ |m| \leq n).
\end{equation}
The eigenvalues are obtained by solving the mixed boundary value
problem (\ref{eq:u_def2}):
\begin{equation}  \label{eq:mu_shell}
\mu_n^{(p)} = g'_n(R)  \qquad (n=0,1,2,\ldots),
\end{equation}
where
\begin{equation}
g_n(r) = \frac{k'_n(\alpha R) i_n(\alpha r) - i'_n(\alpha R) k_n(\alpha r)}
{k'_n(\alpha R) i_n(\alpha L) - i'_n(\alpha R) k_n(\alpha L)} \,,
\end{equation}
$\alpha = \sqrt{p/D}$, and 
\begin{align*}
i_n(z) & = \sqrt{\pi/(2z)}\, I_{n+1/2}(z) , \\
k_n(z) & = \sqrt{2/(\pi z)}\, K_{n+1/2}(z)
\end{align*}
are the modified spherical Bessel functions of the first and second
kind, respectively.  The eigenfunctions do not depend on $p$, whereas
the eigenvalues $\mu_n^{(p)}$ do not depend on the second index $m$
and are thus $(2n+1)$ times degenerate.  In the limit $p\to 0$, one
gets
\begin{equation}  \label{eq:mu_shell_p0}
\mu_n^{(0)} = \frac{n(n+1)}{L} \, \frac{1 - (R/L)^{2n+1}}{n + 1 + n (R/L)^{2n+1}} \,.
\end{equation}

The surface hopping propagator from Eq. (\ref{eq:Sigma_M}) reads
\begin{equation}
\Sigma_p(\s,\ell|\s_0) = \frac{1}{L^2}\sum\limits_{n=0}^\infty \sum\limits_{m=-n}^n Y_{mn}^*(\theta_0,\phi_0) \, 
Y_{mn}(\theta,\phi) \, e^{-\mu_n^{(p)} \ell} . 
\end{equation}
Since the eigenvalues do not depend on the index $m$, one can apply
the addition theorem for spherical harmonics to evaluate the sum over
$m$:
\begin{equation}  \label{eq:Sigma_shell}
\Sigma_p(\s,\ell|\s_0) = \frac{1}{4\pi L^2} \sum\limits_{n=0}^\infty (2n+1) P_n\biggl(\frac{(\s\cdot \s_0)}{|\s|\, |\s_0|}\biggr)  e^{-\mu_n^{(p)}\ell} ,
\end{equation}
where $P_n(z)$ are Legendre polynomials.

One also needs to compute $V_n^{(p)}(\x_0)$ from Eq. (\ref{eq:Vnp}).
Using the summation formulas from \cite{Grebenkov19g}, the
Laplace-transformed quantities $\tilde{G}_\infty(\x,p|\x_0)$ and thus
$\tilde{j}_\infty(\s,p|\x_0)$ for a spherical shell with Dirichlet
boundary condition on the outer sphere and Neumann boundary condition
on the inner sphere read
\begin{align} \nonumber
\tilde{G}_\infty(\x,p|\x_0) &= \sum\limits_{n=0}^\infty \frac{\alpha (2n+1)}{4\pi D}  P_n\biggl(\frac{(\x \cdot \x_0)}{|\x| \, |\x_0|}\biggr)
g_n(r_0)  \\   \label{eq:Ginf_ND_3d}
& \times \bigl[k_n(\alpha L) i_n(\alpha r) - i_n(\alpha L) k_n(\alpha r)\bigr] ,\\  \label{eq:jinf_ND_3d}
\tilde{j}_\infty(\s,p|\x_0) &= \sum\limits_{n=0}^\infty \frac{2n+1}{4\pi L^2} \, P_n\biggl(\frac{(\s \cdot \x_0)}{|\s| \, |\x_0|}\biggr)
g_n(r_0) ,
\end{align}
where $r = |\x|$, $r_0 = |\x_0|$, $R \leq r_0 \leq r \leq L$, and we
used the Wronskian
\begin{equation} \label{eq:Wronskian_3d}
i'_n(z) k_n(z) - k'_n(z) i_n(z) = \frac{1}{z^2} \, .  
\end{equation}
The projection of $\tilde{j}_\infty(\s,p|\x_0)$ onto an eigenfunction
$v_{nm}(\s)$ from Eq. (\ref{eq:vn_shell}) reads then
\begin{equation}
V_{nm}^{(p)}(\x_0) = v_{mn}(\theta_0,\phi_0) \, g_n(r_0) ,
\end{equation}
where $\x_0 = (r_0,\theta_0,\phi_0)$ in spherical coordinates.  The
orthogonality of spherical harmonics reduces Eq. (\ref{eq:U2p}) to
\begin{equation}   \label{eq:U_shell}
\tilde{U}(\ell,p|\x_0) =  g_0(r_0) \, \exp(-\mu_0^{(p)}\ell)  ,
\end{equation}
while the probability density of the reaction time reads
\begin{equation}   \label{eq:Hpsi_shell}
\tilde{H}_q(p|\x_0) = g_0(r_0) \, \frac{1}{1 + \mu_0^{(p)}/q}  \,.
\end{equation}

\subsection{Interior of a ball}

In the limit $R\to 0$, the inner boundary shrinks to a point, and one
gets the solution for the interior of a ball of radius $L$: $\Omega =
\{ \x\in\R^3 ~:~ |\x| < L\}$, with radial functions
\begin{equation}
g_n(r) = \frac{i_n(r\sqrt{p/D})}{i_n(L\sqrt{p/D})} \,.
\end{equation}
The eigenvalues and eigenfunctions of the Dirichlet-to-Neumann
operator are still given by Eqs. (\ref{eq:vn_shell},
\ref{eq:mu_shell}), and other earlier expressions remain valid;
in particular, Eqs. (\ref{eq:Ginf_ND_3d}, \ref{eq:jinf_ND_3d}) are
applicable.  At $p = 0$, the eigenvalues are simply $\mu_n^{(0)} =
n/L$, for which Eq. (\ref{eq:Sigma_shell}) can be evaluated explicitly
using the generating function of Legendre polynomials:
\begin{equation}  \label{eq:Sigma_ball_p0}
\Sigma_0(\s,\ell|\s_0) = \frac{L}{4\pi} \, \frac{1 - e^{-2\ell/L}}{|e^{-\ell/L} \s_0 - \s|^3} \,.
\end{equation} % CHECKED: A_localtime_check1;
This expression coincides with the harmonic measure density on the
sphere when the starting point is $\s_0 e^{-\ell/L}$.  It can also be
written in terms of the angle $\theta$ between vectors $\s_0$ and
$\s$:
\begin{equation}  \label{eq:Sigma_ball_p02}
\Sigma_0(\s,\ell|\s_0) = \frac{1}{4\pi L^2} \, \frac{1 - e^{-2\ell/L}}{[1 - 2e^{-\ell/L} \cos\theta + e^{-2\ell/L}]^{3/2}} \,.
\end{equation}
Figure \ref{fig:Sigma_ball}(a,b) illustrates the behavior of the
surface hopping propagator. 

The orthogonality of spherical harmonics reduces Eq. (\ref{eq:U2}) to
\begin{equation}
\tilde{U}(\ell,p|\x_0) = \frac{R}{r_0} \, \frac{\sinh(r_0\sqrt{p/D})}{\sinh(R\sqrt{p/D})} \, \exp(-\mu_0^{(p)} \ell) ,
\end{equation}
with $\mu_0^{(p)} = \sqrt{p/D}\, \ctanh(R\sqrt{p/D}) - 1/R$.
Similarly, one has
\begin{equation}
\tilde{H}_q(p|\x_0) = \frac{R}{r_0} \frac{\sinh(r_0 \sqrt{p/D})}{\sinh(R \sqrt{p/D})} \, \frac{1}{1 + \mu_0^{(p)}/q} \,,
\end{equation}
from which the inverse Laplace transform yields the standard spectral
expansion for $H_q(t|\x_0)$.

\subsection{Reactive inner sphere}

The analysis for the reactive inner sphere is very similar.  The
eigenfunctions of the Dirichlet-to-Neumann operator associated with
the inner sphere $\Gamma = \{\x\in\R^3 ~:~ |\x| = R\}$ are again the
spherical harmonics but with the prefactor $1/R$ for a proper
normalization:
\begin{equation}  \label{eq:vn_shell2}
v_{nm}(\s) = \frac{1}{R} \, Y_{mn}(\theta,\phi)  \quad (n=0,1,2,\ldots, ~ |m| \leq n).
\end{equation}
The eigenvalues are 
\begin{equation}  \label{eq:mu_shell2}
\mu_n^{(p)} = -g'_n(R)  \qquad (n=0,1,2,\ldots),
\end{equation}
where
\begin{equation}
g_n(r) = \frac{k'_n(\alpha L) i_n(\alpha r) - i'_n(\alpha L) k_n(\alpha r)}
{k'_n(\alpha L) i_n(\alpha R) - i'_n(\alpha L) k_n(\alpha R)} \,.
\end{equation}
In the limit $p\to 0$, one gets
\begin{equation}  \label{eq:mu_shell2_p0}
\mu_n^{(0)} = \frac{n(n+1)}{R} \, \frac{1 - (R/L)^{2n+1}}{n + (n+1) (R/L)^{2n+1}} \,.
\end{equation}
The expression for the surface hopping propagator is almost identical
to Eq. (\ref{eq:Sigma_shell}):
\begin{equation}  \label{eq:Sigma_shell2}
\Sigma_p(\s,\ell|\s_0) = \frac{1}{4\pi R^2} \sum\limits_{n=0}^\infty (2n+1) P_n\biggl(\frac{(\s\cdot \s_0)}{|\s|\, |\s_0|}\biggr)  e^{-\mu_n^{(p)}\ell} .
\end{equation}

The Laplace-transformed propagator $\tilde{G}_\infty(\x,p|\x_0)$ and
thus $\tilde{j}_\infty(\s,p|\x_0)$ for a spherical shell with
Dirichlet boundary condition on the inner sphere and Neumann boundary
condition on the outer sphere read
\begin{align} \nonumber
\tilde{G}_\infty(\x,p|\x_0) &= \sum\limits_{n=0}^\infty \frac{\alpha (2n+1)}{4\pi D}  P_n\biggl(\frac{(\x \cdot \x_0)}{|\x| \, |\x_0|}\biggr)
g_n(r_0)  \\  \label{eq:Ginf_DN_3d}
& \times \bigl[k_n(\alpha R) i_n(\alpha r) - i_n(\alpha R) k_n(\alpha r)\bigr] ,\\  \label{eq:jinf_DN_3d}
\tilde{j}_\infty(\s,p|\x_0) &= \sum\limits_{n=0}^\infty \frac{2n+1}{4\pi R^2} \, P_n\biggl(\frac{(\s \cdot \x_0)}{|\s| \, |\x_0|}\biggr)
g_n(r_0) ,
\end{align}
where $r = |\x|$, $r_0 = |\x_0|$, $R \leq r \leq r_0 \leq L$, from
which
\begin{equation}
V_{nm}^{(p)}(\x_0) = v_{mn}(\theta_0,\phi_0) \, g_n(r_0) .
\end{equation}
These quantities determine the full propagator $P(\x,\ell,t|\x_0)$.

\subsection{Exterior of a ball}

In the limit $L\to\infty$, the outer boundary is pushed away to
infinity, and one deals with diffusion in the exterior of a ball of
radius $R$: $\Omega = \{ \x\in\R^3 ~:~ |\x|> R\}$.  The radial
functions are reduced to
\begin{equation}
g_n(r) = \frac{k_n(r\sqrt{p/D})}{k_n(R\sqrt{p/D})} \,,
\end{equation}
while the eigenvalues and eigenfunctions of the Dirichlet-to-Neumann
operator are still given by Eqs. (\ref{eq:vn_shell2},
\ref{eq:mu_shell2}).  Interestingly, the eigenvalues are just
polynomials of $\sqrt{p/D}$, e.g., $\mu_0^{(p)} = (1 +
R\sqrt{p/D})/R$.  The above expressions are as well applicable; in
particular, Eqs. (\ref{eq:Ginf_DN_3d}, \ref{eq:jinf_DN_3d}) are valid.

At $p = 0$, the eigenvalues in Eq. (\ref{eq:mu_shell2_p0}) are
simplified as $\mu_n^{(0)} = (n+1)/R$, and the surface hopping
propagator can be computed explicitly as
\begin{equation}  
\Sigma_0(\s,\ell|\s_0) = \frac{R \, e^{-\ell/R}}{4\pi} \, \frac{1 - e^{-2\ell/R}}{|e^{-\ell/R} \s_0 - \s|^3} \,.
\end{equation} % CHECKED: A_localtime_check1;
If $L$ is replaced by $R$, this expression coincides with
Eq. (\ref{eq:Sigma_ball_p0}), except for an extra factor $e^{-\ell/R}$
that accounts for the possibility of escaping to infinity.  Figure
\ref{fig:Sigma_ball}(c,d) illustrates the behavior of the surface
hopping propagator.  

The orthogonality of spherical harmonics reduces Eq. (\ref{eq:U2p}) to
\begin{equation}  
\tilde{U}(\ell,p|\x_0) = \frac{R}{r_0} \, \exp\bigl(-(r_0-R+\ell)\sqrt{p/D} - \ell/R\bigr),
\end{equation}
from which the inverse Laplace transform yields
\begin{equation}
U(\ell,t|\x_0) = \frac{R e^{-\ell/R}}{r_0} \, \frac{r_0-R+\ell}{\sqrt{4\pi Dt^3}} e^{-(r_0 - R+\ell)^2/(4Dt)} .
\end{equation}
This is a rare example when the probability density $U(\ell,t|\x_0)$
is found in a simple closed form.  Setting $\ell = 0$, one retrieves
the probability density of the first-passage time for a perfectly
absorbing sphere \cite{Smoluchowski17}.  In turn, the integral
(\ref{eq:Hpsi}) yields the probability density of the first-passage
time to a partially reactive sphere \cite{Collins49,Grebenkov18}
\begin{align}  
H_q(t|\x_0) & = \frac{qD}{r_0} e^{-(r_0-R)^2/(4Dt)} \biggl\{ \frac{R}{\sqrt{\pi Dt}} \\   \nonumber
& - (1 + qR) \erfcx\biggl(\frac{r_0-R}{\sqrt{4Dt}} + (1+qR) \frac{\sqrt{Dt}}{R}\biggr) \biggr\}.
\end{align}

\subsection{Both reactive spheres}

Finally, the analysis for both reactive spheres is the most involved
but very similar to the planar case.  For this reason, we just
reproduce the main formulas adapted to the three-dimensional case.
Here, one employs two families of radial functions,
\begin{align*}
g_{n,1}(r) & = \frac{k_n(\alpha L) i_n(\alpha r) - i_n(\alpha L) k_n(\alpha r)}
{k_n(\alpha L) i_n(\alpha R) - i_n(\alpha L) k_n(\alpha R)} \,, \\
g_{n,2}(r) & = \frac{k_n(\alpha R) i_n(\alpha r) - i_n(\alpha R) k_n(\alpha r)}
{k_n(\alpha R) i_n(\alpha L)  - i_n(\alpha R) k_n(\alpha L)} \,,
\end{align*}
which satisfy $g_{n,1}(R) = 1$, $g_{n,1}(L) = 0$ and $g_{n,2}(L) = 1$,
$g_{n,2}(R) = 0$.  The eigenfunctions are searched in the form
\begin{equation}
v_{nm}^{(p)}(\s) = \left\{ \begin{array}{l l} a_n^{(p)} \, Y_{mn}(\theta,\phi) & \quad \s \in \Gamma_1 , \\  
b_n^{(p)} \, Y_{mn}(\theta,\phi)  & \quad \s \in \Gamma_2 , \\  \end{array} \right. 
\end{equation}
where $\Gamma_1$ and $\Gamma_2$ are the inner and the outer spheres
forming the boundary $\pa$.  The Wronskian (\ref{eq:Wronskian_3d})
implies
\begin{equation}
g'_{n,1}(L) = \frac{A}{\alpha L^2} \,, \qquad g'_{n,2}(R) = - \frac{A}{\alpha R^2}\, ,
\end{equation}
where
\begin{equation}
A = \frac{1}{k_n(\alpha L) i_n(\alpha R) - i_n(\alpha L) k_n(\alpha R)} \,.
\end{equation}
In this case, one also has
\begin{align}  \nonumber
B & = - g'_{n,1}(R) g'_{n,2}(L) + g'_{n,1}(L) g'_{n,2}(R) \\
& = - \alpha^2 \frac{k'_n(\alpha L) i'_n(\alpha R) - i'_n(\alpha L) k'_n(\alpha R)}{k_n(\alpha L) i_n(\alpha R) - i_n(\alpha L) k_n(\alpha R)} \,.
\end{align}
Using these expressions, one deduces again Eqs. (\ref{eq:mu_shell_DD},
\ref{eq:ab_shell_DD}) for the eigenvalue $\mu_{n,\pm}^{(p)}$ and the
coefficients $a_{n,\pm}^{(p)}$ and $b_{n,\pm}^{(p)}$.  The
normalization coefficient $c_{n,\pm}^{(p)}$ is fixed by normalization:
\begin{align*}  
1 & = [c_{n,+}^{(p)}]^2 \left(4\pi R^2 [g'_{n,2}(L) - \mu_{n,+}^{(p)}]^2 + 4\pi L^2 [g'_{n,1}(L)]^2 \right) , \\
1 & = [c_{n,-}^{(p)}]^2 \left(4\pi R^2  [g'_{n,2}(R)]^2 + 4\pi L^2 [g'_{n,1}(R) + \mu_{n,-}^{(p)}]^2\right) .
\end{align*}

In the limit $p\to 0$, one gets
\begin{align*}
g'_{n,1}(R) & \to - \frac{n+1 + n(R/L)^{2n+1}}{R(1 - (R/L)^{2n+1})} \,,\\
g'_{n,1}(L) & \to -  \frac{(2n+1) (R/L)^{n+1}}{L(1 - (R/L)^{2n+1})} \,,\\
g'_{n,2}(R) & \to \frac{(2n+1) (R/L)^n}{R(1 - (R/L)^{2n+1})} \,,\\
g'_{n,2}(L) & \to \frac{n + (n+1)(R/L)^{2n+1}}{L(1 - (R/L)^{2n+1})} \,,
\end{align*}
(and $B \to n(n+1)/(LR)$), from which
\begin{align}
& \mu_{n,\pm}^{(0)} = \frac{1}{2R(1-\gamma_n)} \biggl(n(\beta+1)+1 + \gamma_n(n+(n+1)\beta) \\  \nonumber
& \pm \sqrt{(n(\beta+1)+1 + \gamma_n(n+(n+1)\beta))^2 - 4\beta n(n+1)} \biggr) ,
\end{align}
where $\beta = R/L$ and $\gamma_n = (R/L)^{2n+1}$.  We also get
\begin{equation}
\mu_{0,-}^{(0)} = 0,  \qquad \mu_{0,+}^{(0)} = \frac{1 + \beta^2}{R(1-\beta)} 
\end{equation}
and
\begin{align*}
a_{0,-}^{(0)} & = b_{0,-}^{(0)} = \frac{1}{\sqrt{4\pi(R^2+L^2)}} \,, \\
a_{0,+}^{(0)} & = \frac{1}{R\sqrt{4\pi (1+\beta^2)}} , \qquad b_{0,+}^{(0)} = \frac{-\beta^2}{R \sqrt{4\pi (1 + \beta^2)}}  \,.
\end{align*}
In the limit $L\to \infty$, one retrieves $\mu_{n,+}^{(0)} \to
(n+1)/R$ and $\mu_{n,-}^{(0)} \to 0$.  In turn, as $R\to 0$, one has
$\mu_{n,-}^{(0)} \to n/L$, whereas $\mu_{n,+}^{(0)} \to \infty$.

Using the asymptotic behavior of the modified spherical Bessel
functions, one can also check that
\begin{align*}
\lim\limits_{R\to 0} \mu_{n,-}^{(p)} & = \sqrt{p/D} \,\frac{i'_n(L\sqrt{p/D})}{i_n(L\sqrt{p/D})} \,, \\
\lim\limits_{R\to 0} \mu_{n,+}^{(p)} & = \infty \,, \\
\lim\limits_{L\to \infty} \mu_{n,+}^{(p)} & = - \sqrt{p/D} \,\frac{k'_n(R\sqrt{p/D})}{k_n(R\sqrt{p/D})} \,, \\
\lim\limits_{L\to \infty} \mu_{n,-}^{(p)} & = \sqrt{p/D}  \,,
\end{align*}
where we used that $k'_n(z)/k_n(z) \leq - 1$.  As a consequence, in
the limit $R\to 0$, one retrieves the eigenvalues of $\M_p$ for the
interior of a ball of radius $L$.  In turn, in the limit $L\to
\infty$, $\mu_{n,+}^{(p)}$ approach the eigenvalues for the exterior
of a ball of radius $R$, while $\mu_{n,-}^{(p)}$ accumulate near
$\sqrt{p/D}$.

Finally, the Dirichlet propagator in the Laplace domain is
\begin{subequations}
\begin{align}   \nonumber
& \tilde{G}_\infty(\x,p|\x_0) = \sum\limits_{n=0}^\infty \frac{\alpha (2n+1)}{4\pi D}  P_n\biggl(\frac{(\x \cdot \x_0)}{|\x| \, |\x_0|}\biggr)
 \, g_{n,2}(r_0) \\   
& \times \bigr[k_n(\alpha L) i_n(\alpha r) - i_n(\alpha L) k_n(\alpha r)\bigr] \quad  (r_0 \leq r),  \\   \nonumber
& \hskip 19mm = \sum\limits_{n=0}^\infty \frac{\alpha (2n+1)}{4\pi D}  P_n\biggl(\frac{(\x \cdot \x_0)}{|\x| \, |\x_0|}\biggr)
 \, g_{n,1}(r_0) \\   
& \times \bigr[k_n(\alpha R) i_n(\alpha r) - i_n(\alpha R) k_n(\alpha r)\bigr] \quad  (r \leq r_0),
\end{align}
\end{subequations}
where $\x = (r,\theta,\phi)$ and $\x_0 = (r_0,\theta_0,\phi_0)$.  The
probability flux density reads then
\begin{subequations}
\begin{align}  \nonumber
& \tilde{j}_\infty(\s,p|\x_0) \\
&= \sum\limits_{n=0}^\infty \frac{2n+1}{4\pi R^2}  
P_n\biggl(\frac{(\x \cdot \x_0)}{|\x| \, |\x_0|}\biggr) \, g_{n,1}(r_0)  \quad (\s\in\Gamma_1) , \\
&= \sum\limits_{n=0}^\infty \frac{2n+1}{4\pi R^2}  
P_n\biggl(\frac{(\x \cdot \x_0)}{|\x| \, |\x_0|}\biggr) \, g_{n,2}(r_0)  \quad (\s\in\Gamma_2) .
\end{align}
\end{subequations}
As a consequence, one gets
\begin{equation}
V_n^{(p)}(\x_0) = \bigl(a_n^{(p)} \, g_{n,1}(r_0) + b_n^{(p)} \, g_{n,2}(r_0) \bigr) Y_{mn}(\theta_0,\phi_0)
\end{equation}
that gives access to the full propagator $P(\x,\ell,t|\x_0)$.

\end{document}